\documentclass[onecolumn]{article}
\usepackage[a4paper,margin=1.8cm]{geometry}
\usepackage{cuted}
\usepackage[caption=false, font=footnotesize]{subfig}
\usepackage{amsmath}
\usepackage{amssymb}
\usepackage{amsfonts}
\usepackage{booktabs}
\usepackage{colortbl}
\usepackage{graphicx}
\usepackage{multirow}
\usepackage{tikz}
\usetikzlibrary{shapes.geometric, arrows, calc}
\usepackage{tabularx}
\usepackage{placeins}

\definecolor{light-blue}{rgb}{0.8,0.85,1}
\definecolor{light-gray}{rgb}{0.83, 0.83, 0.83}
\definecolor{pastel-red}{rgb}{1.0, 0.41, 0.38}
\definecolor{pastel-yellow}{rgb}{0.99, 0.99, 0.59}
\definecolor{pastel-green}{rgb}{0.47, 0.87, 0.47}

\tikzstyle{block} = [rectangle, rounded corners, minimum width=4cm, minimum height=0.6cm,text centered, draw=black]
\tikzstyle{operation} = [circle, minimum size = 0.7cm, text centered, draw = black]
\tikzstyle{arrow} = [thick,->,>=stealth]

\title{Spectral Segmented Linear Regression for  Coarse Carrier Frequency Offset Estimation in Optical LEO Satellite Communications}

\author{I.~P.~Vieira,~G.~V.~Serra,~R.~A.~Colares,~D.~A.~A.~Mello%
	\thanks{IPV, GVS, RAC, and DAAM are with the Department of Communications (DECOM), School of Electrical and Computer Engineering, University of Campinas (UNICAMP), Campinas, SP, 13083-852, BR (email: darli@unicamp.br).}
	\thanks{This work was supported by \textit{the Conselho Nacional de Desenvolvimento Cient\'{i}fico e Tecnol\'{o}gico (CNPq)} and by \textit{the Fapesp AltaVe Multiuser Equipment Project} \#2022/11596-0.}
}%
\date{}%

\begin{document}
	
\maketitle
	
\begin{abstract}
		Carrier frequency offset estimation (CFOE) is a critical stage in modern coherent optical communication systems. Although conventional all-digital techniques perform reliably in typical fiber-optic communication links, CFOE can become a major bottleneck in low-symbol-rate scenarios with large carrier frequency offsets (CFOs) approaching the signal bandwidth and severe additive noise levels. These conditions are particularly prevalent in links between optical ground stations (OGSs) and low Earth orbit (LEO) satellites, where Doppler-induced frequency shifts of several gigahertz and atmospheric attenuation can significantly degrade CFOE performance and can render conventional methods ineffective. In this paper, we propose a robust non-data-aided (NDA) scheme designed for wide-range CFOE. The proposed coarse CFOE (C-CFOE) algorithm partially compensates the CFO, enabling the operation of a subsequent fine CFOE stage. By applying low-complexity operations to the spectrum of the received signal, we recast the frequency estimation task as a segmented linear regression (SLR) problem. Numerical simulations in stress-test scenarios involving large CFOs, low SNR, and low symbol rates show that the proposed approach achieves good estimation accuracy and robust convergence. Offline experimental validation further confirms the practical feasibility of the method. 
\end{abstract}

\section{Introduction}
Coherent optical detection combined with advanced digital signal processing (DSP) techniques has become a core technology of modern optical communications \cite{hui2025introduction}. This paradigm leads to significant improvements in spectral efficiency and receiver sensitivity enabled by techniques such as multilevel modulation formats and all-digital compensation of transmission impairments. These capabilities have steered the evolution of industry standards toward ultra-high per-wavelength data rates approaching the terabit scale in fiber-optic communication systems \cite{OIF800ZR2024}. In free-space optical (FSO) systems, experimental demonstrations have achieved data rates ranging from tens of gigabits per second (Gbps), using intensity-modulation/direct-detection (IM/DD) techniques \cite{song20044times10,chen2008demonstration}, to on the order of 100~Gbps with coherent detection, including long-reach links involving satellite communications \cite{walsh2022demonstration}. These data rates are expected to increase substantially in the coming years, potentially reaching the terabit-per-second regime and beyond \cite{ellis2023fso,boddeda2024current}. As FSO systems progressively adopt DSP-based coherent receivers to enable higher capacities \cite{bitachon2022tbit,schieler2023orbit}, accurate carrier synchronization becomes a key performance-limiting factor.

Links established between optical ground stations (OGSs) and low-Earth-orbit (LEO) satellites represent some of the most challenging operating scenarios, owing to a combination of severe physical-layer impairments and stringent system constraints. The architecture of present-day ultra-dense LEO constellations \cite{pachler2024flooding}, with satellites traveling at a typical orbital velocity of 7.8~km per second \cite{ESA_LEO_2020}, can give rise to Doppler-induced frequency offsets on the order of 4~GHz at a wavelength of 1550~nm \cite{fernandes2023digitally}. Moreover, the absence of in-line optical amplification, together with propagation impairments such as scattering, absorption, and atmospheric turbulence, forces operation under degraded signal-to-noise ratio (SNR) conditions. In practice, maintaining connectivity in distance-adaptive links may require reducing the symbol rate, which further increases phase noise-related impairments \cite{mello2021digital}. These factors collectively challenge conventional CFO estimation (CFOE) techniques developed for fiber-based systems, motivating the need for robust, wide-range, and low-complexity synchronization algorithms tailored to coherent optical LEO satellite communications. Frequency-domain CFOE techniques exploiting spectral features, such as spectral edges, spectral asymmetry, or power differences over predefined frequency regions, have been previously investigated, particularly in systems where the signal position within the receiver bandwidth can be inferred from the shape of the received spectrum \cite{diniz2011simple,liu2014improved,zhang2010dual,yi2013frequency}. However, these approaches typically rely on local spectral descriptors or scalar metrics extracted from selected portions of the PSD.  In the operating regime considered in this work, characterized by large CFOs, reduced symbol rates, and low SNR, such local features may become noisy, weakly defined, or sensitive to the choice of frequency regions. Therefore, a coarse CFOE strategy that exploits the global structure of the received spectrum is desirable.

In this paper, we propose a novel non-data-aided (NDA) CFOE algorithm based on segmented linear regression (SLR) applied to the accumulated power spectral density (PSD) of the received signal. Because the proposed estimator operates on the accumulated PSD rather than on constellation-dependent decisions, its principle is not tied to a specific modulation format. For Nyquist-shaped signals, the spectral transitions exploited by the method remain available for uniform and non-uniform constellations, including higher-order QAM formats and probabilistic-constellation-shaped signals. Differently from edge-based or asymmetry-based spectral methods, the proposed approach treats the accumulated PSD as a structured piecewise linear signal and estimates the CFO from the joint location of two transition points that delimit the signal-occupied region within the receiver bandwidth. Hence, the CFO is not inferred from a single spectral edge, a scalar asymmetry metric, or the relative power measured over predefined frequency bands, but from the global shape of the accumulated spectrum. The cumulative representation also provides an integration-induced smoothing effect, which improves robustness under low-SNR conditions, while the SLR formulation offers a systematic and low-complexity procedure to locate the spectral transitions.

The algorithm operates as a coarse CFOE (C-CFOE) stage that partially estimates and compensates the CFO, enabling the operation of subsequent fine CFOE (F-CFOE) algorithms applied to the residual CFO. The rationale behind the method is that, ideally, the accumulated PSD of a Nyquist-shaped signal detected with excess receiver bandwidth is the concatenation of three approximately linear segments: two with similar slopes, corresponding to the left and right noise-floor regions, and one with a steeper slope, corresponding to the signal-occupied portion of the spectrum. Such excess receiver bandwidth is expected in distance-adaptive satellite links, where symbol rates are reduced to support sustained connectivity under adverse attenuation conditions. The method is validated by simulation and laboratory experiments under stringent requirements of near-zero SNR, low symbol rates, and high frequency offsets, as often encountered in LEO satellite connections under challenging atmospheric and Doppler-shift conditions.

The remainder of this paper is organized as follows. Section \ref{sec:works} compiles related works. Section \ref{sec:methodology} introduces the proposed method, detailing its theoretical foundations and implementation. Section \ref{sec:results} presents the simulation and experimental results. Finally, Section \ref{sec:conclusion} concludes the paper.
	
\section{Related Works}
\label{sec:works}

Several techniques have been proposed to address large CFOs in optical communication systems. In the context of satellite communications, Doppler-shift (DS) estimation is particularly demanding, as estimators may need to cope with offsets ranging from hundreds of megahertz to a few gigahertz \cite{yang2009doppler,vieira2023modulation,fernandes2023digitally}. State-of-the-art wide-range solutions include adaptive recursive estimators, such as Kalman filters \cite{xu2025robust}, for tracking time-varying offsets based on channel models, and optical phase-locked loops (OPLLs) and optical injection locking (OIL) for hardware-assisted carrier recovery \cite{rosenkranz2016receiver,shoji2012pilot,ren2022adaptive}. In fully digital implementations, the techniques range from pilot- or training-aided schemes to fully NDA approaches operating in either the time or frequency domain \cite{diniz2011simple,pita2022all,fernandes2023digitally}.

State-space recursive estimators have been extensively used to model and correct linear signal distortions induced by the communication channel. In a recent study, Xu \textit{et al.} \cite{xu2025robust} propose a robust two-stage Kalman-filter (TS-KF) scheme to compensate both abrupt and gradual DS variations during the establishment and maintenance of LEO-LEO optical inter-satellite links (OISLs). Using a 60-GBd polarization-division-multiplexed QPSK coherent optical simulation system with a cosmic channel, the authors show that the proposed TS-KF scheme outperforms the conventional Viterbi-Viterbi (VV) and blind phase search (BPS) algorithms in terms of bit error rate (BER). In particular, the scheme compensates abrupt DS values up to 10~GHz while tracking gradual DS drift rates of 151.73~MHz/s at an OSNR of 18~dB, maintaining the BER below the FEC threshold of $3.8\times10^{-3}$.

Among optical-domain carrier-recovery techniques, OPLLs stand out as one of the most established approaches. Liu \textit{et al.} propose in \cite{liu2018study} a multistage composite OPLL for DS compensation and carrier recovery in satellite coherent optical communication systems. The scheme combines temperature tuning, piezoelectric (PZT) tuning, and acousto-optic frequency shifter (AOFS) tuning in a compound Costas-loop architecture, aiming to reconcile a wide frequency-acquisition range with high-speed phase locking. The outer temperature-tuning loop performs coarse Doppler compensation over a wide range, while the PZT and AOFS inner loops provide residual frequency correction and fine phase locking. In the experimental validation, a BPSK signal at 5~Gbps is used to emulate inter-satellite DSs from 0 to 4~GHz. The proposed OPLL achieves a lock range of 4~GHz, a loop bandwidth of 1.7~MHz, and a residual phase error of approximately $5.1^\circ$, enabling data recovery with a BER of $1\times10^{-7}$ at a received optical power of $-41.2$~dBm. Rosenkranz and Schaefer present in \cite{rosenkranz2016receiver} a comparative numerical study of carrier recovery strategies for OISLs, investigating homodyne detection based on OPLLs and benchmarking it against intradyne detection assisted by digital CFO and phase noise compensation. In the homodyne architecture, the OPLL adjusts the local-oscillator (LO) frequency and phase to match the incoming signal, but its hardware complexity increases with the modulation order. As an alternative, the authors consider a DSP-based intradyne receiver, in which C-CFO compensation is performed using a phase-differential algorithm, followed by F-CFO compensation based on the VV algorithm. For a simulated BPSK/QPSK OISL system operating at 1~Gbaud with a residual intradyne frequency offset of 100~MHz, the DSP-based intradyne approach achieved nearly the same receiver sensitivity as the homodyne OPLL-based scheme, indicating the potential of digital CFO compensation for reducing receiver complexity in OISL systems.

Some fully digital approaches rely on transmitter-side parameter adaptation, assuming prior knowledge or prediction of the DS evolution. For example, Almonacil \textit{et al.} propose in \cite{almonacil2020digital} an ephemeris-aided transmitter-side digital DS compensation scheme for satellite optical communications, combining digital pre-emphasis (DPE) and signal clipping to alleviate DAC bandwidth limitations. For a 64-Gbaud 16-QAM signal, the method enables compensation of Doppler shifts up to 10~GHz at a 500-Gbps line rate, without laser wavelength tuning, while requiring less than 0.7~dB of additional optical launch power. In~\cite{fernandes2023digitally}, Fernandes \textit{et al.} introduce a fully digital DS mitigation framework for coherent LEO-to-Earth FSO links based on dynamic symbol-rate adaptation combined with probabilistic constellation shaping (PCS). By jointly adapting the symbol rate and constellation entropy, the method significantly extends the tolerable CFO range, experimentally supporting offsets exceeding $\pm 15~\mathrm{GHz}$ in a 600~Gbps transmission.

All-digital data-aided (DA) schemes for CFOE have also been extensively investigated. Zhou \textit{et al.} propose in ~\cite{zhou2011wide} a training-sequence-based wide-range CFOE algorithm that removes the modulated phase information by exploiting known training symbols, rather than relying on the traditional $M$-th power operation. By decoupling the frequency estimation process from the modulation format, the method achieves an estimation range approaching half of the symbol rate, $\pm R_s/2$, independently of the modulation order, and demonstrates estimation errors on the order of a few megahertz for frequency offsets of several gigahertz in simulated 28-GBaud QPSK, 8-PSK, and 16-QAM systems. Similarly, Zhao \textit{et al.}  introduce in  \cite{zhao2015digital} a digital pilot-aided CFOE technique in which a specially designed pilot sequence generates a pilot-tone-like spectral component, whose frequency shift can be tracked in the frequency domain. Owing to this spectral structure, the method achieves a wide estimation range theoretically bounded by the receiver sampling rate and exhibits high estimation accuracy, with sub-megahertz variance under moderate optical SNR (OSNR) conditions. In addition, the approach is shown to be largely insensitive to residual chromatic dispersion and polarization-mode dispersion, and to operate independently of the modulation format. He \textit{et al.} propose in ~\cite{he2025almost} an almost-blind wide-range CFOE framework in the context of discrete-spectrum nonlinear frequency-division multiplexing (DS-NFDM) systems. By combining an eigenvalue-shift-based coarse estimation stage with a refined phase-error minimization criterion, the method achieves wide CFO estimation ranges without baud-rate constraints, while requiring only two training symbols to resolve periodic ambiguities. Experimental results demonstrate accurate CFO estimation for offsets on the order of several hundreds of megahertz, with reported training-symbol overhead reductions exceeding 94-99\% compared to fully training-assisted approaches.

Although DA schemes can provide accurate and wide-range CFO estimates, their reliance on known training symbols introduces overhead and may reduce spectral efficiency. To avoid this limitation, NDA CFOE techniques have been investigated as an alternative, estimating the frequency offset directly from the received data signal without requiring pilot or training information. In~\cite{pita2022all}, we present a two-stage compensation strategy combining a C-CFOE method derived from~\cite{diniz2011simple} with a subsequent $M$-th power-based fine correction, targeting DS mitigation in optical inter-satellite links (OISLs) under architectures representative of recent commercial LEO constellations. The C-CFOE stage, originally proposed by Diniz \textit{et al.} in \cite{diniz2011simple}, is based on an imbalance between the positive ($P_{+}$) and negative ($P_{-}$) spectral components of the received signal, where the CFO is estimated as being proportional to the logarithm of the ratio $P_{+}/P_{-}$. Chino \textit{et al.} propose in \cite{chino2025two} an interesting two-stage CFO estimation method for inverse scattering transform (IST)-based transmission systems, where eigenvalue shifts enable wide-range acquisition and the scattering coefficient $b$ is exploited for fine estimation. While the approach achieves wide estimation range and high accuracy, it is inherently tied to nonlinear Fourier transform (NFT)/IST signal representations and cannot be directly applied to conventional linear coherent modulation formats.

Other CFO estimation methods have been proposed for spectrally efficient transmission architectures, such as digital subcarrier multiplexing (DSM) and discrete spectrum \cite{yi2013frequency,liu2014improved,zhang2010dual,xing2005frequency}. These methods typically exploit spectral features, such as spectral edges, dips, or power imbalance, to infer the frequency offset without explicit training symbols. Overall, existing CFO estimation techniques face a trade-off between estimation range, accuracy, computational complexity, and reliance on training overhead. In particular, the problem of wide-range NDA CFOE under low symbol-rate and low-SNR conditions remains insufficiently addressed -- as illustrated in Table \ref{tab:methods} --, motivating the development of alternative approaches that combine wide acquisition range with robust convergence and low implementation complexity.

\begin{table*}[ht]
    \centering
    \caption{Comparison of wide-range CFO/DS estimation and compensation methods for coherent optical communication systems. SIA stands for side-information-aided.}
    \resizebox{\textwidth}{!}{%
    \begin{tabular}{|c|c|c|c|c|c|c|}
        \hline
        Method & Approach & Domain & Link scenario & Min. $R_s$/Modulation & Max. $\Delta f$ & Min. SNR/OSNR/Pwr. \\
        \hline
        \cite{xu2025robust} 
        & NDA
        & Time
        & LEO-LEO 
        & 60 GBd/PDM-QPSK 
        & 10 GHz 
        & OSNR $\approx$ 18 dB \\
        \hline
        \cite{liu2018study}
        & OPLL
        & Optical
        & MEO-MEO
        & 5 Gb/s BPSK  ($\approx$5 GBd)
        & 4 GHz
        & $P_{\mathrm{Rx}}=-41.2$ dBm \\
        \hline
        \cite{shoji2012pilot}
        & DA/OIL
        & Optical
        & LEO-OGS
        & 10 Gb/s BPSK ($\approx$10 GBd)
        & 10.3 GHz
        & OSNR = 9.7 dB \\
        \hline
        \cite{almonacil2020digital} 
        & SIA 
        & Frequency
        & LEO-LEO 
        & 32 GBd/16QAM 
        & 10 GHz 
        & OSNR $\approx$ 15 dB \\
        \hline
        \cite{fernandes2023digitally}
        & SIA
        & Tx-DSP
        & LEO-OGS
        & 65 GBd/PCS-64QAM
        & $15$ GHz
        & SNR $\approx 15$ dB \\
        \hline
        \cite{diniz2011simple}
        & NDA
        & Frequency
        & Fiber
        & 28 GBd/PM-QPSK
        & $6$ GHz
        & OSNR $\approx 15$ dB \\
        \hline
        \cite{liu2014improved}
        & NDA
        & Frequency
        & Fiber
        & 20 GBd/QPSK
        & $2.4$ GHz
        & SNR $\approx 7$ dB \\
        \hline
        \cite{zhang2010dual}
        & NDA
        & Time/frequency
        & Fiber
        & 10.7 GBd/PM-QPSK
        & 9 GHz
        & OSNR = 6 dB \\
        \hline
    \end{tabular}%
    }
    \label{tab:methods}
\end{table*} 
	
\section{Methodology}
\label{sec:methodology}
\subsection{Principle of Operation}

The proposed CFOE algorithm takes the periodogram of the oversampled received signal as a starting point. The sampling frequency, $F_s$, must be such that it allows the maximum expected frequency offset, $\Delta f_{max}$, to be adequately accommodated in the receiver bandwidth. This is a reasonable assumption in distance-adaptive satellite links, where the symbol rate can be strongly reduced to operate at stringent attenuation conditions. Next, the signal's periodogram PSD is cumulated over frequency throughout the entire reception band. For a discrete-time signal $y[n]$, $n = 0, 1, \cdots, N-1$, with DC-centered discrete Fourier transform (DFT) $Y[k] = \left(\sum_{n=0}^{N-1}y[n]e^{-j2\pi (k-N/2)n/N}\right)$, $k = 0,1,\cdots,N-1$, the cumulative bandlimited power reads:
\begin{equation}
P_{\text{acc}}(f_m) = \frac{1}{NF_s}\sum_{k=0}^{m: f_k \leq f_m} \left| Y[k] \right|^2 \delta f, \quad f_m \in \left[-F_s/2,F_s/2\right),
\label{eq:P_acc}
\end{equation}
where $\delta f = F_s/N$ is the frequency resolution and $f_k = \left(k-N/2\right)\delta f$ is the frequency value at the $k$-th bin.

\begin{figure}[ht]
    \centering
    \includegraphics[width=\textwidth]{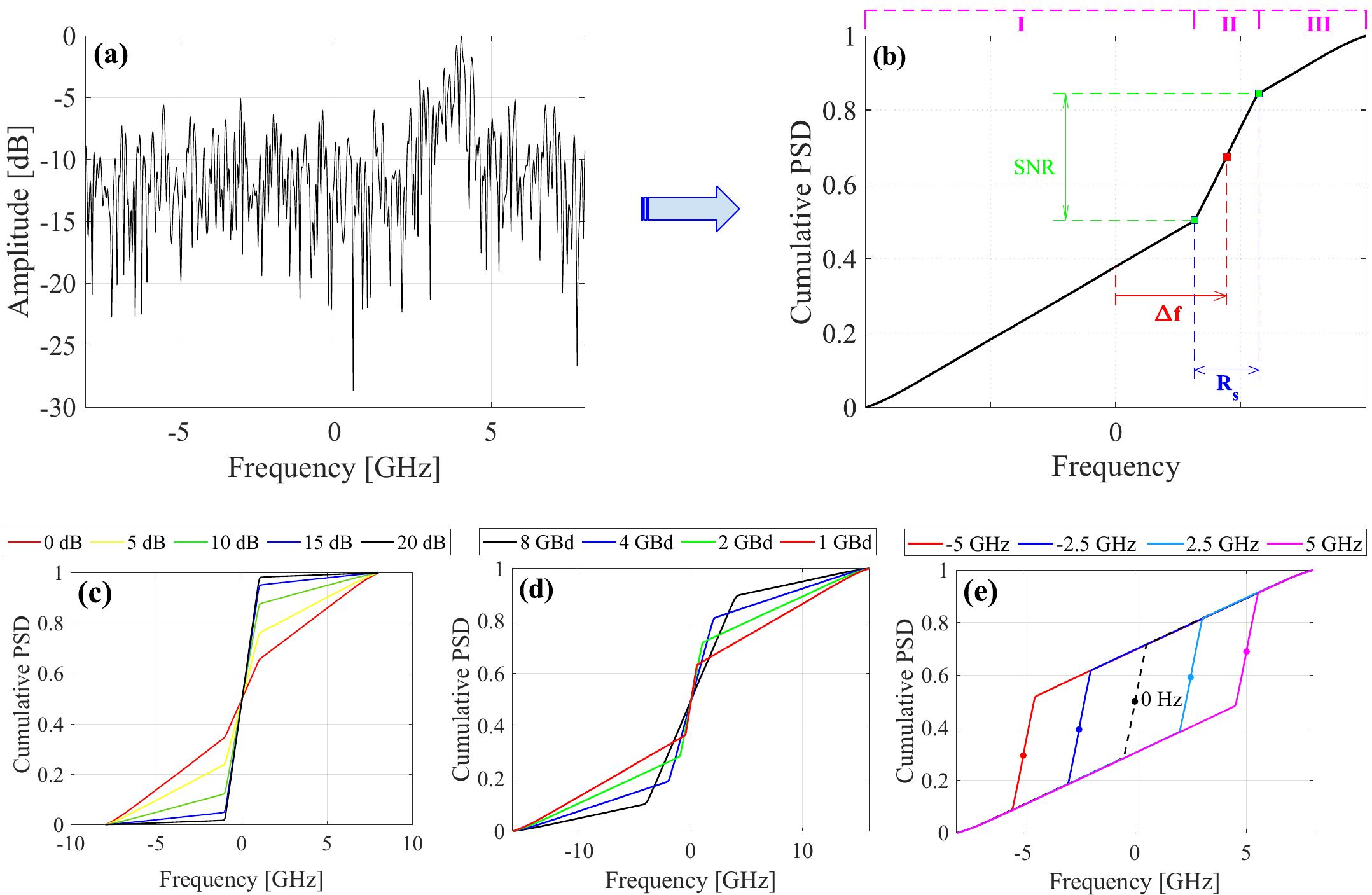}
    \caption{(a) Spectrum of a 2-GBd PM-QPSK signal, shaped by a root-raised-cosine (RRC) filter with roll-off factor $\alpha = 0.1$, at 1-dB SNR per bit and impaired by a 3.5-GHz CFO and its (b) corresponding normalized cumulative linear PSD -- with SNR, signal bandwidth ($R_s(1+\alpha)$), and CFO ($\Delta f$) indicated --, forming a characteristic three-segment piecewise-linear structure. Impact of different (c) per bit SNR levels, (d) symbol rates, and (e) CFOs in the cumulative PSD behavior. The accumulation process suppresses high-frequency fluctuations, yielding a more pronounced contrast across the noise-floor and signal-region regimes under SNR variation.}
    \label{fig:spec2segreg}
\end{figure}

\indent\indent Figure \ref{fig:spec2segreg}a depicts the PSD of a 2-GBd polarization-multiplexed (PM)-QPSK signal impaired by a frequency offset of $\Delta f = 3.5\textrm{ GHz}$, computed using a 1024-symbol FFT block. Applying cumulative integration to the PSD yields the characteristic profile shown in Fig. \ref{fig:spec2segreg}b. Line segments I and III correspond to the result of the integration process in the noise floor region, thus having slopes that are close to each other and lower than that of segment II, referring to the region of the spectrum that contains the signal. In this sense, the projection of segment II onto the $f$-axis represents the signal's bandwidth. Accordingly, in baseband, the distance from its midpoint to the origin yields an estimate of the frequency offset to which the signal is subjected.

 \subsection{Segmented linear regression (SLR)}

 In this paper, we propose the use of SLR techniques to estimate the CFO of the received signal, shaping and discriminating regions I, II and III in the spectral domain. SLR models assume that the conditional mean function of a response variable $y$ given a predictor $x$, $\displaystyle E[y\mid x]$, is piecewise linear, showing distinct linear relationships over specific intervals of the covariate space. Formally, let $\Psi = \left\{\psi_1<\psi_2<\cdots<\psi_K\right\}$ be an ordered set of structural change points, referred to as breakpoints, that partition the domain of $x$ into $K+1$ contiguous segments such that
\begin{equation}
E[y \mid x] =
\begin{cases}
\beta_{0,1} + \beta_{1,1} x, & x < \psi_1, \\
\beta_{0,2} + \beta_{1,2} x, & \psi_1 \le x < \psi_2, \\
\vdots \\
\beta_{0,K+1} + \beta_{1,K+1} x, & x \ge \psi_K,
\end{cases}
\end{equation}
where $B = \left\{\left(\beta_{0,j},\beta_{1,j}\right)\right\}_{j = 1}^{K+1}$ denotes the collection of regime-specific intercept and slope parameters. In general, both the regression parameters and the breakpoints are unknown and must be estimated from the data, most often by minimizing a least-squares criterion.

Under the continuity constraint, $\beta_{0,j}+\beta_{1,j}\psi_j = \beta_{0,j+1}+\beta_{1,j+1}\psi_{j}\textrm{, }\forall j = 1,\cdots,K$, the conditional mean function $E[y \mid x]$ is allowed to change slope but not level, and the model simplifies to the canonical compact parametric representation \cite{muggeo2003estimating}
\begin{equation}
    y_i = \beta_0+\beta_1x_i + \sum_{k=1}^K\varphi_k\left(x_i-\psi_k\right)_++\varepsilon_i,\quad \psi_1<\psi_2<\cdots<\psi_K,
\end{equation}
where $\beta_0$ ($\beta_1$) is the baseline intercept (slope) -- i.e., that applying for $x<\psi_1$ --, $\varphi_k$ corresponds to the incremental slope contribution after the $k$-th breakpoint, $\left(\cdot\right)_+ = \textrm{max}\left\{0,\cdot\right\}$ is the hinge function, and $\varepsilon_i$ is a random error term satisfying $E[\varepsilon_i \mid x_i] = 0$ and $\textrm{Var}\left(\varepsilon \mid x_i\right) = \sigma^2 < \infty$.

The successful SLR application depends on three basic assumptions   \cite{seber2003nonlinear}: (i) the number of regimes is small relative to the sample size; (ii) within each regime, the conditional expectation is well approximated by linear functions; (iii) transitions between adjacent regimes are abrupt -- manifested as changes in slope -- rather than gradual. Figures \ref{fig:spec2segreg}c-e show the result of the PSD accumulation process over a 1024-sample FFT block for different SNR levels, symbol rates, and CFOs, respectively.

Assumptions (i) and (ii) are directly related to the intrinsic spectral structure of the received signal. For a band-limited Nyquist-shaped signal, the PSD exhibits, in first approximation, a step-like profile composed of three main regions: the noise floor before the signal band, the occupied signal band, and the noise floor after the signal band. Therefore, the number of regimes is naturally small compared with the number of FFT samples, satisfying assumption (i). Moreover, since the noise floor outside the signal band is approximately flat and the in-band PSD is approximately bounded by a reasonably defined spectral plateau, the accumulated PSD tends to produce nearly piecewise-linear trends over these regions. As a result, within each regime, the conditional expectation can be well approximated by a linear function, supporting assumption (ii). Assumption (iii), in turn, tends to become less robust as the SNR decreases, since lower SNR levels smooth out the abrupt boundaries between adjacent regimes. Nevertheless, a clear violation arises only in the limiting case where the SNR approaches zero.

\subsection{Algorithm}
\label{subsec:algorithm}

\begin{figure}[tb]
    \centering
    \includegraphics[width=\linewidth]{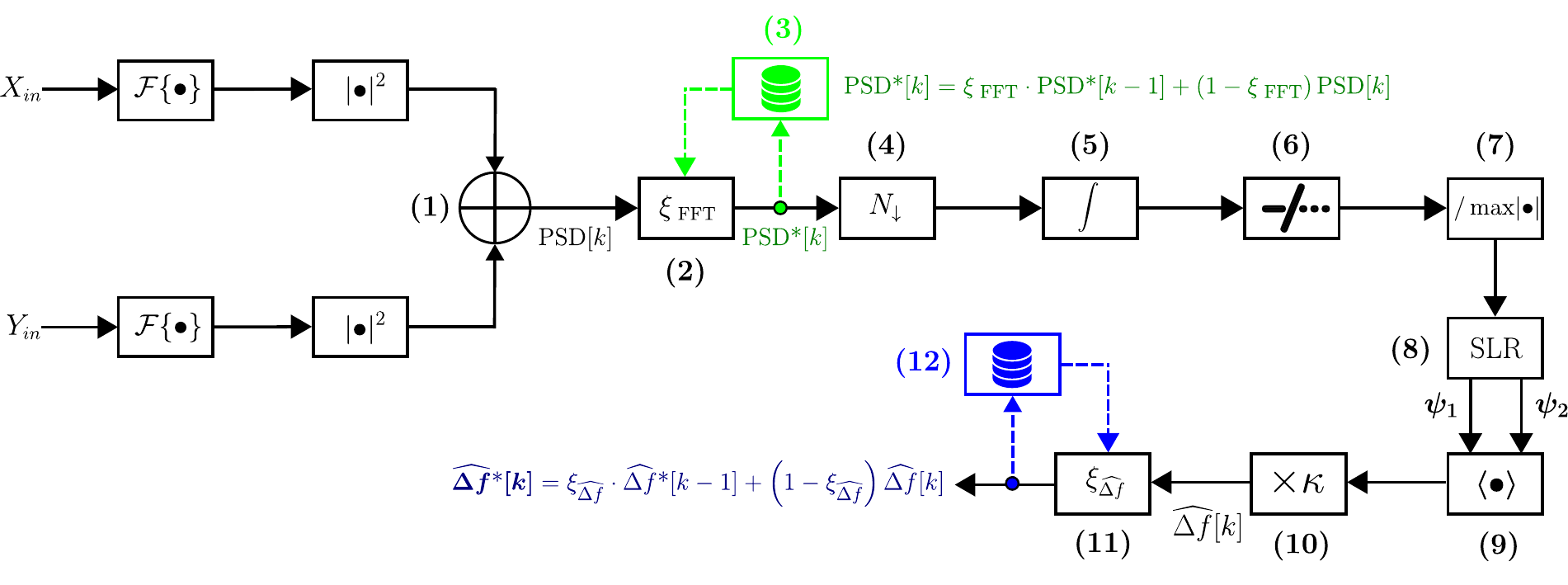}
    \caption{Schematic of the C-CFOE method. The received samples from the two polarizations are first processed by block-wise FFTs and magnitude-squared operations to obtain the corresponding PSD estimates, which are then combined in (1). The combined PSD is smoothed using the forgetting factor $\xi_{\mathrm{FFT}}$ in (2), with the previous smoothed PSD stored in (3). The resulting spectrum is downsampled by $N_{\downarrow}$ in (4), accumulated in (5), trimmed at the spectral boundaries in (6), and normalized by its maximum absolute value in (7). The normalized accumulated PSD is then processed by the SLR stage in (8), which estimates the breakpoints $\psi_1$ and $\psi_2$. Their midpoint is computed in (9) and rescaled to physical frequency units by the factor $\kappa$ in (10), yielding the block-wise CFO estimate $\widehat{\Delta f}[k]$. Finally, the estimate is temporally smoothed in (11) using $\xi_{\widehat{\Delta f}}$, with the previous smoothed estimate stored in (12), producing the final coarse CFO estimate $\widehat{\Delta f}^{*}[k]$.}
    \label{fig:diagram}
\end{figure}

This section presents the general concepts underlying the spectral SLR-based C-CFOE. Figure ~\ref{fig:diagram} shows the complete flow diagram of the proposed C-CFOE method, which is applied to each FFT block, including all processing stages leading to the frequency offset estimation. As can be seen, both before and after the application of the SLR method, some processing steps are required to ensure the quality of the estimation. In what follows, we conduct a walkthrough of the diagram blocks, discussing their roles in the methodology.  

\subsubsection{Data Preprocessing and Numerical Conditioning}

Before the CFOE can be effectively performed through the SLR algorithm, a set of preprocessing steps must be carried out to ensure reliable operation and numerical stability. The DSP flow begins by splitting the received signal, initially in the time domain, into fixed-size blocks and computing their corresponding frequency-domain representations using the DFT. Subsequently, the PSD of each block is calculated for both polarizations. Since the frequency offset affecting each polarization is approximately the same, the two PSDs are combined, contributing to the reduction of the relative noise power. At this stage, a forgetting factor $\xi_{\text{FFT}}$ is applied to suppress high-frequency noise fluctuations, resulting in more stable FFT estimates. Following this, a downsampling step is performed to remove surplus samples produced by the analog-to-digital converter (ADC). The excess sampling requirement arises from two main factors. First, the receiver must accommodate the maximum expected frequency offset, $\Delta f_{\max}$, to which the signal may be subjected. Second, due to the way the method is constructed, a minimum spectral margin must be ensured to properly define the three operating regimes -- where the signal region is effectively confined between the noise floors. Accordingly, the downsampling factor is expressed as
\begin{equation}
N_{\downarrow} = 2^{\left\lfloor\log_2\!\left(\frac{F_s}{2 \max\!\left\{\frac{R_s(1+\alpha)}{2} + \Delta f_{\max},\;
R_s\right\}}\right)\right\rfloor},
\label{eq:downsample}
\end{equation}
where $\lfloor\cdot\rfloor$ denotes the floor operator, $\max\{\cdot,\cdot\}$ returns the maximum of its two arguments, and $\alpha$ is the roll-off factor of the pulse-shaping filter. Equation~(\ref{eq:downsample}) has a direct spectral interpretation. After downsampling, the available Nyquist frequency range is limited to $\pm F_s/(2N_{\downarrow})$. Therefore, this range must be sufficiently wide to contain the pulse-shaped signal spectrum under the worst-case CFO condition. The term $R_s(1+\alpha)/2$ represents the spectral half-width of the pulse-shaped signal around its own center frequency, while $\Delta f_{\max}$ accounts for the maximum translation of this spectrum within the receiver bandwidth. Hence, $R_s(1+\alpha)/2+\Delta f_{\max}$ corresponds to the maximum spectral excursion that must be accommodated after downsampling, avoiding spectral truncation of the shifted signal spectrum. The second term inside the maximum operator, $R_s$, prevents the downsampling factor from becoming excessively large when the CFO requirement is less restrictive -- in fact, it enforces a post-downsampling sampling rate not lower than $2R_s$, ensuring a sufficiently resolved spectral representation of the signal for the subsequent DSP stages. Increasing the roll-off factor enlarges the excess bandwidth of the pulse-shaped signal and reduces the maximum allowable downsampling factor. The next step involves integrating the filtered PSD across the entire reception band, as described in Eq. (\ref{eq:P_acc}). 

\begin{figure}[ht!]
    \centering
    \includegraphics[width=0.65\textwidth]{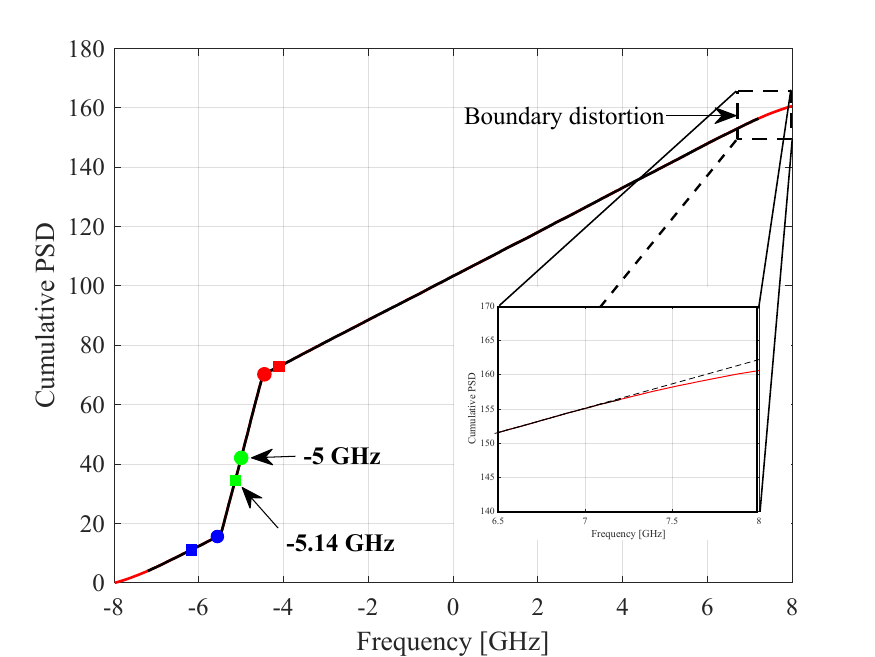}
    \caption{Impact of FFT boundary-bin exclusion on frequency-estimation accuracy, emphasizing distortions introduced by pulse shaping and finite observation windows. Spectral-edge artifacts arise when the PSD is accumulated for a RRC-shaped 1-GBd PM-QPSK signal (span: 20 symbols, roll-off factor $\alpha = 0.1$), at 64 samples per symbol and impaired by a -5 GHz CFO. The blue and red squares mark the breakpoints $\psi_1$ and $\psi_2$, respectively, obtained from the full PSD and frequency vectors, whereas the circles indicate the estimates after excluding edge samples. For the considered FFT block, using the complete cumulative PSD and frequency vectors results in a CFO overestimation of approximately -140 MHz (green square). In contrast, removing boundary bins mitigates edge distortions and yields an accurate CFO estimate, as indicated by the green circle.}
    \label{fig:boundary1}
\end{figure}

Due to the finite observation window and pulse shaping transients at the block boundaries, FFT bins located near the edges of the occupied spectrum typically exhibit reduced and distorted spectral content, particularly at low symbol rates, as shown in Fig. \ref{fig:boundary1}. To mitigate this effect and ensure reliable CFOE, frequency bins corresponding to the spectral boundaries are excluded from the estimation process.

Following the diagram flow, the next stage corresponds to axis normalization: the accumulated PSD is scaled such that its maximum value equals unity, and the frequency axis ranges $\left[-0.5, 0.5\right]$. At this point, the samples are ready to proceed to the SLR algorithm, where the two breakpoints, $\psi_1$ and $\psi_2$, are estimated. It follows that the average of the breakpoints multiplied by the normalization scale factor results in the estimation of the block frequency offset, $\widehat{\Delta f}$. Finally, to enhance the stability of the estimation, a new forgetting factor, $\xi_{\widehat{\Delta f}}$, is introduced within $\widehat{\Delta f}$, implementing a low-pass filtering operation.

\subsubsection{Breakpoint and Slope Estimation}

We now summarize the closed-form, implementation-ready procedure used to estimate the two breakpoints, $\psi_1$ and $\psi_2$, as well as the corresponding slopes of a continuous three-segment piecewise-linear model, following the approach originally proposed by Jacquelin~\cite{Jacquelin2009}. The derivation of the formulas employed here is provided in Appendix~\ref{appA}.

Let $\{(x_i,y_i)\}_{i=1}^{N}$ be the input data, where $x_i$ denotes the independent variable and $y_i$ is the response, with strictly increasing abscissae, $x_1<\cdots<x_N$. The goal is to fit a continuous piecewise-linear function with slopes $p \equiv \left(p_1, p_2, p_3\right)^T$, separated by two unknown breakpoints $\psi_1<\psi_2$, and intercepts $q \equiv\left(q_1, q_2, q_3\right)^T$ for each segment. The estimation proceeds in two stages: first, an auxiliary regression is used to estimate $\psi_1,\psi_2$ (steps 1 to 4); second, a standard linear regression is used to estimate the slopes and intercepts (steps 5 and 6).

\medskip
\noindent\textbf{Step 1 (Discrete primitives).}
Define $S_y[i]$ and $S_{xy}[i]$ as discrete numerical primitives of the sampled quantities $y$ and $xy$, respectively, computed by the trapezoidal rule:
\[
S_y[1]=0,\qquad 
S_y[i]=S_y[i-1]+\frac{y_{i-1}+y_i}{2}\,(x_i-x_{i-1}),
\]
\[
S_{xy}[1]=0,\qquad 
S_{xy}[i]=S_{xy}[i-1]+\frac{x_{i-1}y_{i-1}+x_i y_i}{2}\,(x_i-x_{i-1}),
\qquad i=2,\ldots,N.
\]
Thus, $S_y[i]\approx\int_{x_1}^{x_i}y(x)dx$ and $S_{xy}[i]\approx\int_{x_1}^{x_i}xy(x)dx$. 

\medskip
\noindent\textbf{Step 2 (Construction of breakpoint regressors).}
Using the quantities above, construct the auxiliary regressors. For each $i=1,\ldots,N$, define
\[
\begin{aligned}
F_{0,i} &= y_i,\\
F_{1,i} &= 6S_{xy}[i]-2x_iS_y[i]-x_i^2y_i,\\
F_{2,i} &= x_i y_i-2S_y[i],\\
F_{3,i} &= x_i,\\
F_{4,i} &= 1.
\end{aligned}
\]
Here, $\mathbf F_i\equiv(F_{1,i},F_{2,i},F_{3,i},F_{4,i})^{\mathsf T}$ denotes the predictor vector of an auxiliary linear regression whose coefficients are used to compute the breakpoint locations.

\medskip
\noindent\textbf{Step 3 (Estimation of auxiliary coefficients).} Form the normal equation matrix, $M_F$, and the corresponding right-hand-side vector, $b_F$, of this regression as
\[
M_F=\sum_{i=1}^{N}\mathbf F_i\mathbf F_i^{\mathsf T},
\qquad
b_F=\sum_{i=1}^{N}F_{0,i}\mathbf F_i.
\]
The coefficient vector $C \equiv (C_1,C_2,C_3,C_4)^{\mathsf T}$ is then obtained by solving the linear system
\[
C=M_F^{-1}b_F.
\]

\medskip
\noindent\textbf{Step 4 (Closed-form breakpoint estimation).}
The breakpoints are given as the two roots of a quadratic expression involving $C_1$ and $C_2$:
\[
\psi_{1,2}=\frac{C_2\mp\sqrt{C_2^{\,2}-4C_1}}{2C_1},
\qquad \psi_1<\psi_2.
\]

\medskip
\noindent\textbf{Step 5 (Construction of slope regressors).}
Given $\psi_1$ and $\psi_2$, we construct regressors that represent a continuous piecewise-linear function with three segments. Let $H(t)=\mathbf 1_{t\ge0}$ be the Heaviside function. For each $i = 1, \cdots, N$, define
\[
\begin{aligned}
G_{1,i} &= x_i-(x_i-\psi_1)H(x_i-\psi_1),\\
G_{2,i} &= (x_i-\psi_1)H(x_i-\psi_1)-(x_i-\psi_2)H(x_i-\psi_2),\\
G_{3,i} &= (x_i-\psi_2)H(x_i-\psi_2),\\
G_{4,i} &= 1.
\end{aligned}
\]
These regressors $\mathbf G_i\equiv(G_{1,i},G_{2,i},G_{3,i},G_{4,i})^{\mathsf T}$ define the contributions of the first, second, and third linear segments over the intervals $x<\psi_1\textrm{, }\psi_1\leq x < \psi_2\textrm{, and }x\geq\psi_2$, respectively, while preserving continuity of the fitted function.

\medskip
\noindent\textbf{Step 6 (Estimation of slopes).} Form the normal equation matrix, $M_G$, and right-hand-side vector, $b_G$, of this regression as
\[
M_G=\sum_{i=1}^{N}\mathbf G_i\mathbf G_i^{\mathsf T},
\qquad
b_G=\sum_{i=1}^{N}y_i\,\mathbf G_i,
\]
and solve
\[
(p_1,p_2,p_3,q_1)^{\mathsf T}=M_G^{-1}b_G.
\]
Here, $p_1,p_2\textrm{, and }p_3$ determine the segment gradients, while $q_1$ corresponds to the intercept of the first line segment. The continuity at 
$\psi_1$ and $\psi_2$ uniquely determines the remaining intercepts $q_2$ and $q_3$ as
\[
q_2 = q_1 + (p_1 - p_2)\psi_1,
\qquad
q_3 = q_2 + (p_2 - p_3)\psi_2.
\]

\medskip
\noindent

The overall procedure requires computing two cumulative sums and the solution of two linear systems of size $4\times4$, with no nonlinear optimization or iterative search. Furthermore, it is worth noting that, for the CFO estimation purposes, simply obtaining the breakpoints in step 4 is sufficient.

\begin{figure}[ht]
    \centering
    \includegraphics[width=0.65\linewidth]{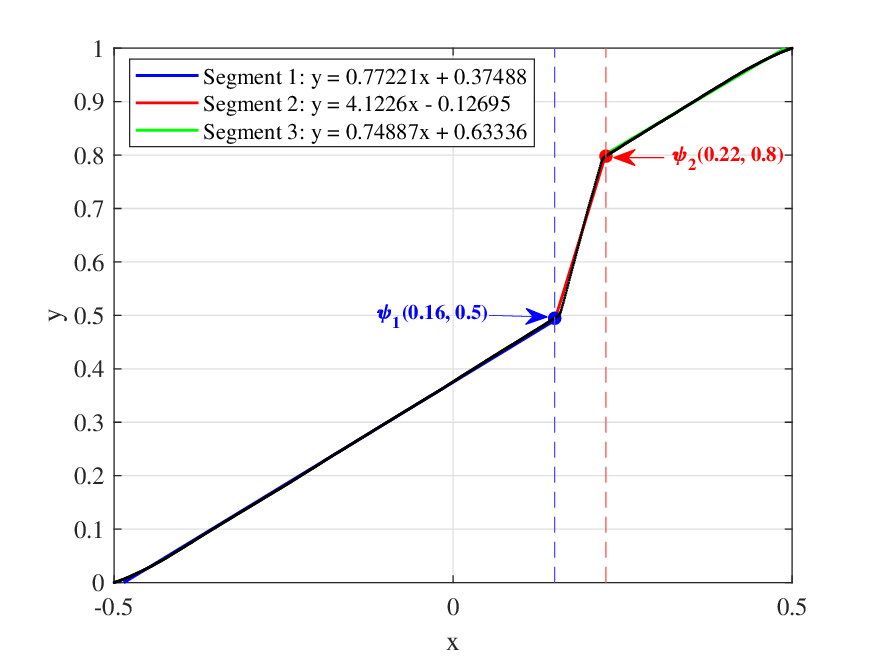}
    \caption{SLR-based estimation applied to a 1-GBd QPSK signal affected by a 3-GHz CFO. Both axes are normalized to enhance numerical stability.}
    \label{fig:estimation}
\end{figure}

Figure~\ref{fig:estimation} illustrates the application of the method to a dual-polarization QPSK transmission at a symbol rate of 1~GBd, under a frequency offset three times larger than the symbol rate. The estimation corresponds to the final FFT block of the signal, after both the FFT processing and the estimator have fully stabilized under the influence of the forgetting factors $\xi_{\textrm{FFT}}$ and $\xi_{\widehat{\Delta f}}$. Normalization of the $x$- and $y$-axes improves numerical stability. Two breakpoints, $\psi_1 = (0.16,\,0.5)$ and $\psi_2 = (0.22,\,0.8)$, are identified, along with the slopes of the three corresponding line segments.

\subsection{Complexity}

The SLR procedure for breakpoint estimation requires $\sim60N$ elementary arithmetic operations, where $N$ denotes the number of data samples, as summarized in Tab.~\ref{tab:elementary_operations}. Consequently, the computational complexity of the SLR procedure follows $\mathcal{O}(N)$\footnote{It should be noted that the FFT required to obtain the frequency-domain representation of the received signal has complexity $\mathcal{O}(N\log N)$. However, FFT-based processing is common to several frequency-domain DSP blocks in coherent optical receivers and, once the signal is represented in the frequency domain, this representation can be reused across the receiver chain. If the FFT is accounted for as an exclusive operation of the proposed estimator, the overall complexity scales with $\mathcal{O}(N\log N)$, while the estimator-specific post-FFT processing remains linear in $N$.}. From a hardware implementation perspective, the Jacquelin method is well-suited for real-time signal processing owing to its non-iterative structure and linear computational complexity. The algorithm follows a fixed sequence of elementary arithmetic operations without requiring iterative refinement, matrix factorizations, or transcendental function evaluations. This deterministic execution flow ensures predictable latency and bounded computational cost, which are critical for deployment in high-throughput, low-latency optical signal processing systems. Furthermore, the streaming data access pattern requires only constant memory and enables straightforward pipelining, making the method particularly amenable to efficient fixed-point implementations on embedded processors.

\noindent\begin{table}[th]
\centering
\caption{Number of elementary arithmetic operations (additions/subtractions, multiplications, and divisions) required at each step of the Jacquelin SLR algorithm. The operation counts assume that the inverses of the $4\times4$ matrices involved in the normal equations are precomputed, so that solving the linear systems reduces to a matrix-vector multiplication.}
\label{tab:elementary_operations}
\begin{tabularx}{\linewidth}{l l c c l}
\hline
\textbf{Step} & \textbf{Add/Sub} & \textbf{Mult} & \textbf{Div} & \textbf{Remark}\\
\hline\hline
\textbf{1}& $5\left(N-1\right)$ & $7\left(N-1\right)$ & 0 & Trapezoidal-rule accumulation of $S_y$ and $S_{xy}$ \\
\textbf{2} & $3N$ & $7N$  & 0  & Construction of $\mathbf{F_i}$ \\
\textbf{3} & $4\left(5N+2\right)$ & $4\left(5N+4\right)$  & 0 & Formation of $M_F$, $b_F$, and matrix-vector product \\
\textbf{4} & 3 & 3 & 2 & $\psi_1$ and $\psi_2$; \textbf{one square-root evaluation}\\
\textbf{5} & 6$N$ & 4$N$ & 0 & Construction of $\mathbf{G_i}$ \\
\textbf{6} & $4\left(5N+2\right)$+6 & $4\left(5N+4\right)$  & 0 & Formation of $M_G$, $b_G$, and matrix-vector product \\
\hline
\end{tabularx}
\end{table}

Beyond Jacquelin's formulation of SLR, alternative strategies can also be considered. In particular, methods based on iterative refinement and likelihood-based criteria -- such as those implemented in the \textit{segmented} \cite{muggeo2008segmented} and \textit{piecewise-regression} \cite{pilgrim2021piecewise} packages -- provide a more flexible formulation for breakpoint estimation. These methods require initial guesses for the breakpoint locations, which are subsequently refined through an iterative optimization procedure (e.g., via score-based or Newton-type updates), allowing for the joint estimation of both regression parameters and breakpoint locations. As iterative procedures, however, they may be sensitive to initialization and are not guaranteed to converge in all cases; moreover, they are generally more computationally demanding than the closed-form approach considered here. For this reason, such strategies are not pursued in this work, as the focus is on low-complexity implementations with predictable computational cost.

\section{Results and Discussion}
\label{sec:results}

This section evaluates the performance of the proposed SLR-based CFOE strategy under operating conditions representative of optical satellite communications, assuming a dual-polarization QPSK transmission system throughout the analysis. Numerical simulations are first used to assess estimation range, accuracy, and convergence behavior in the presence of large CFOs, low SNRs, and reduced symbol rates (corresponding to increased phase-noise impact). The robustness of the proposed approach is further examined through offline experimental validation, based on long-duration signal captures, demonstrating frequency-drift tracking capability and characterizing the residual frequency offsets obtained from real measurement data. The results presented here are obtained using Jacquelin’s integral-equation-based linearization approach for segmented linear regression \cite{Jacquelin2009}. Further methodological details are given in Appendix~\ref{appA}.

\subsection{Numerical Simulations}

In practical optical systems, transmitter and LO lasers are subject to systematic frequency deviations as well as time-varying fluctuations resulting from imperfect calibration, thermal drift, environmental perturbations, and intrinsic phase noise. Accordingly, the instantaneous carrier frequency deviation is typically modeled with a sinusoidal profile
\begin{equation}
\Delta f(t) = f_{\textrm{mean}} + \frac{f_{\mathrm{pk\text{-}pk}}}{2}\sin\!\left(2\pi f_j t\right),
\label{eq:freq}
\end{equation}
where $f_{\textrm{mean}}$ is the mean CFO,  $f_{\mathrm{pk\text{-}pk}}$ denotes the peak-to-peak frequency excursion, $f_j$ is the frequency of the imposed temporal variation, and $t$ is the time.

\begin{figure}[t]
    \centering
    \includegraphics[width = 0.65\textwidth]{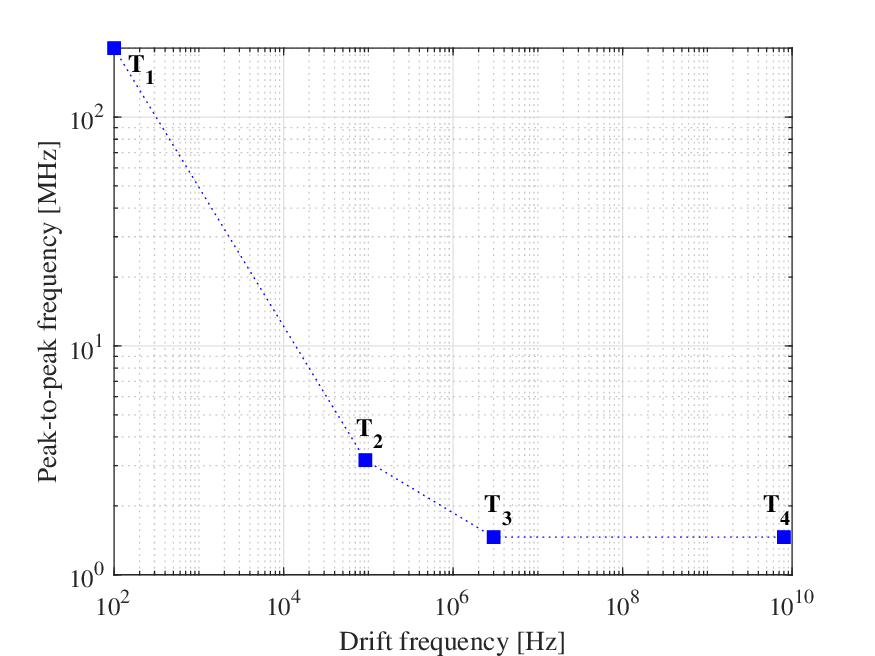}
    \caption{Combinations of $f_{\mathrm{pk\text{-}pk}}$ (y-axis) and $f_j$ (x-axis) that produce four different tones (T$_1$–T$_4$), following the description in \cite{OIF800ZR2024}.}
    \label{fig:oif800}
\end{figure}

The OIF-800ZR standard~\cite{OIF800ZR2024} specifies four distinct frequency tones $\Delta f(t)$. These tones, illustrated in Fig.~\ref{fig:oif800}, define the parameters $f_{\mathrm{pk\text{-}pk}}$ (vertical axis) and $f_j$ (horizontal axis) in Eq. (\ref{eq:freq}). Although they originate from the coherent fiber-optic communication context and should not be interpreted as a physical Doppler model of LEO satellite optical links, they provide standardized and reproducible dynamic CFO profiles. Their use is motivated by the fact that coherent optical satellite communication systems reuse and adapt several receiver-side principles originally developed for coherent fiber-optic systems, including DSP-based compensation stages. In addition, the Doppler-induced frequency offset in satellite optical links evolves much more slowly than the symbol-rate timescale, so that its dominant contribution can be treated as quasi-static over the short observation window used for coarse CFO estimation. It follows that the OIF-800ZR tones are adopted here as a stress-test benchmark for controlled dynamic frequency-offset conditions, while the large mean CFO values are independently chosen to emulate the satellite-link frequency-offset regime.

In the first simulation round, we apply these four tones in three extreme operating conditions (scenarios (a) to (c)) representative of optical satellite links, which are independently assessed by varying one stress condition at a time. The numerical values implemented in the three scenarios are presented in Table \ref{tab:simulation_scenarios}. Scenario~(a) corresponds to a large $f_{\textrm{mean}}$, as typically encountered in LEO-LEO inter-satellite links \cite{vieira2023modulation}. Scenario~(b) addresses low signal-to-noise ratio (SNR) conditions, which are characteristic of links traversing the Earth’s atmosphere, such as LEO-OGS or OGS-LEO links \cite{zhu2002free,andrews2005laser}. Finally, scenario~(c) considers an operation at a low symbol rate, which can occur in the context of distance-adaptive transmission \cite{chatzidiamantis2011adaptive,djordjevic2009communication}. 

\begin{table}[t]
\centering
\caption{Simulation Scenarios.}
\label{tab:simulation_scenarios}
\begin{tabular}{c l c c c}
\hline
\textbf{Scenario} & \textbf{Description} & $(f_{\mathrm{mean}})_{\max}$ (GHz) & SNR (dB) & $R_s$ (GBd) \\
\hline
(a) & Large CFO        & 10 & 15 & 32 \\
(b) & Low SNR          & 5  & 0  & 32 \\
(c) & Low Symbol Rate  & 1  & 15 & 4  \\
\hline
\end{tabular}
\end{table}

Table~\ref{tab:stress} presents the estimation error results for the combination of three scenarios and four $\Delta f(t)$ tones. Each entry reports the worst estimation error per block relative to the applied reference CFO, i.e., $\max \lvert \Delta f[k]-\widehat{\Delta f}^{*}[k]\rvert$ for the $k$-th block, evaluated over 50 independent channel realizations. For each configuration, the mean CFO parameter $f_{\mathrm{mean}}$ in Eq. \eqref{eq:freq} is randomly drawn from a uniform distribution whose support is bounded by the maximum absolute value of the mean CFO, denoted by $(f_{\mathrm{mean}})_{\max}$, such that $\lvert f_{\mathrm{mean}}\rvert \leq (f_{\mathrm{mean}})_{\max}$\footnote{For instance, when $(f_{\mathrm{mean}})_{\max}=10$~GHz, the realizations are generated with $f_{\mathrm{mean}} \sim \mathcal{U}\!\left([-10~\mathrm{GHz},\,10~\mathrm{GHz}]\right)$. In the algorithm, while $f_{\mathrm{mean}}$ spans multiple values within this interval across realizations, the maximum expected CFO is kept fixed as $\Delta f_{\max} = (f_{\mathrm{mean}})_{\max} + f_{\mathrm{pk\text{-}pk}}/2$.}. The random values are generated using a Mersenne Twister pseudo-random number generator initialized with seed 0. When the proposed estimator is employed as a first-stage CFO recovery (i.e., a C-CFOR), the resulting residual frequency offsets consistently lie within the correction range of the 4th-power algorithm ($R_s/8$), thereby enabling accurate fine CFO compensation in a subsequent stage. Moreover, the residual frequency estimates remain largely consistent across the different applied tones, indicating that the estimator performance is only weakly dependent on the specific tone characteristics. In the large-offset regime (scenario~a), the CFO is reduced from 10~GHz to sub-gigahertz residuals, with worst-case values below 800~MHz. Although some dispersion is observed across tones -- most notably for T$_4$ -- the residual offsets remain more than a factor of four below the upper correction limit of the 4th-power algorithm. Under low-SNR operation (scenario~b), the estimator exhibits remarkably stable behavior, yielding nearly identical residual offsets across all tones. Despite the elevated noise level, which increases convergence time, the residual CFO remains around 1.7~GHz, demonstrating that estimator performance is largely insensitive to tone-dependent dynamics in noise-dominated regimes. Finally, in the low-symbol-rate case (scenario~c), corresponding to increased phase-noise impact, the estimator achieves residual offsets on the order of 60~MHz -- more than an order of magnitude reduction relative to the initial offset and well below the F-CFOE threshold. The narrow spread across tones further indicates stable behavior across the different tones even when the symbol rate is reduced to one quarter of the CFO.

\begin{table}[t]
\centering
\caption{Maximum CFO estimation error evaluated across 50 realizations under three stress-case conditions: (a) Large CFO ($\Delta f = 10$ GHz, $\mathrm{SNR} = 15$ dB, $R_s = 32$ GBd); (b) Low SNR ($\Delta f = 5$ GHz, $\mathrm{SNR} = 0$ dB, $R_s = 32$ GBd); and (c) Low symbol rate ($\Delta f = 1$ GHz, $\mathrm{SNR} = 15$ dB, $R_s = 4$ GBd). In each case, the first $100$ FFT blocks were discarded to account for algorithm convergence.}
\label{tab:stress}
\begin{tabular}{c c c c c}
\hline
\textbf{Scenario} & \textbf{Tone 1 (T$_1$)} & \textbf{Tone 2 (T$_2$)} & \textbf{Tone 3 (T$_3$)} & \textbf{Tone 4 (T$_4$)} \\
\hline
(a) & 521.25 MHz & 521.05 MHz & 449.13 MHz & 763.82 MHz \\
(b) & 1.69 GHz   & 1.69 GHz   & 1.68 GHz   & 1.69 GHz   \\
(c) & 57.04 MHz  & 56.26 MHz  & 57.72 MHz  & 57.67 MHz  \\
\hline
\end{tabular}
\end{table}
Taken together, these results indicate that, for the scenarios considered, the proposed estimator delivers consistent and predictable residual CFO levels across a wide range of operating conditions, with only weak dependence on the specific tone characteristics. Next, this analysis is extended to more stringent conditions in which these impairments are jointly applied, allowing the estimator performance to be assessed under compounded stress.

\begin{figure}[ht]
    \centering
    \includegraphics[width=\linewidth]{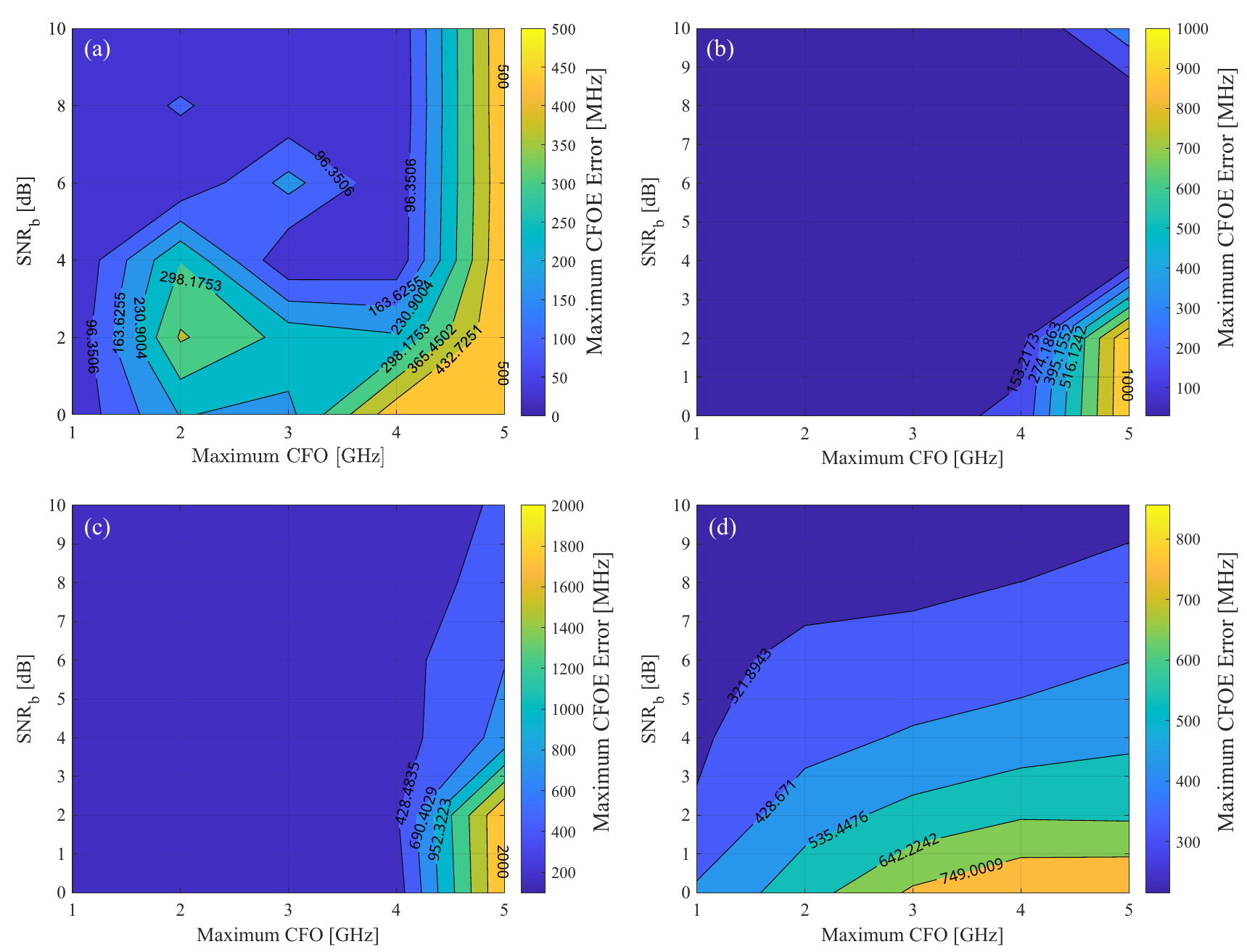}
    \caption{The estimator performance is evaluated under multiple operating scenarios, considering 100 independent channel realizations and a total of $2^{18}$ transmitted symbols. The maximum estimation error is defined as the largest deviation from the reference frequency observed after 100 FFT blocks. Operating regimes combining low SNR and low symbol rate with large frequency offsets constitute the most challenging cases from a carrier synchronization perspective. Using FFTs of 1024 samples in the estimator and setting $\xi_{\mathrm{FFT}} = \xi_{\widehat{\Delta f}} = 0.98$, the proposed method is able to produce a residual frequency offset smaller than $R_s/8$, which is required by the 4th-power algorithm for F-CFOE, across all evaluated scenarios with $\Delta f_{\max} \lesssim 4~\mathrm{GHz}$, even at $\mathrm{SNR}_b = 1~\mathrm{dB}$. In Fig.~(a), corresponding to $R_s = 4~\mathrm{GBd}$, the allowable $\Delta f_{\max}$ margin gradually increases with SNR. In Figs.~(b) and (c), with $R_s = 8~\mathrm{GBd}$ and $R_s = 16~\mathrm{GBd}$, respectively, the estimator operates reliably for $\Delta f_{\max} \leq 5~\mathrm{GHz}$ over the entire region where $\mathrm{SNR}_b > 2.5~\mathrm{dB}$. Finally, in Fig.~(d), corresponding to $R_s = 32~\mathrm{GBd}$, the estimator remains operational over the full evaluated region, maintaining a comfortable margin with respect to the maximum allowable residual frequency offset of $4~\mathrm{GHz}$.
    }
	\label{fig:heatmap}
\end{figure}

Figure~\ref{fig:heatmap} presents the maximum estimation error of the coarse frequency estimator for four symbol rates -- namely 4, 8, 16, and 32~GBd -- while jointly varying the bit-wise SNR (SNR$_b$) between 0 and 10~dB and the maximum value for the frequency offset, $\Delta f(t)$, between 1 and 5~GHz. In our implementation, the receiver operates at a fixed sampling rate of $F_s = \mathrm{64~GSa/s}$, spanning $R_s = 2^{n}$~GBd ($2 \le n \le 5$). For the CFO, we consider the same sinusoidal profile in Eq. (\ref{eq:freq}), with $f_\textrm{mean}\in\mathcal{U}\left(-\Delta f_{\max},\Delta f_{\max}\right)$, $f_{\mathrm{pk\text{-}pk}} = 200$~MHz, and $f_j = 100$ kHz. As expected, the most challenging operating regime across all cases corresponds to the simultaneous presence of large CFOs and low SNR$_b$, where noise-induced estimation uncertainty and reduced observation quality -- stemming from the finite resolution of the FFT -- hinder reliable CFO estimation. Despite this compounded impairment, the estimator exhibits robust performance over a broad region of the parameter space. In particular, for $\Delta f < 4$~GHz and SNR$_b > 1$~dB, the C-CFOE consistently reduces the residual frequency offset to within the capture range required by the subsequent 4th-power algorithm. This behavior is observed across all evaluated symbol rates, indicating that the proposed estimator maintains effective acquisition capability even as the symbol rate decreases and the relative CFO becomes more severe, given that the 4th-power algorithm exhibits a baud-rate-dependent capture range ($\Delta f = \pm R_s/8$). Overall, these results demonstrate that reliable coarse CFO correction can be achieved under stringent operating conditions relevant to optical satellite communication systems, thereby providing a solid foundation for subsequent fine synchronization stages. It is worth emphasizing that Fig.~\ref{fig:heatmap} provides an estimator-level characterization of the failure boundary. From the standpoint of the SLR formulation discussed in Sec. \ref{subsec:algorithm}, estimator degradation occurs when the accumulated PSD no longer exhibits sufficiently distinguishable piecewise-linear regimes. Therefore, the breakdown region can be directly identified in the heatmaps as the set of operating points for which the residual CFO after C-CFOE exceeds the capture range required by the subsequent 4th-power F-CFOE stage. This criterion isolates the behavior of the proposed coarse estimator, avoiding the ambiguity of attributing end-to-end receiver failures to other DSP blocks.

\begin{figure}[ht]
    \centering
    \includegraphics[width=\linewidth]{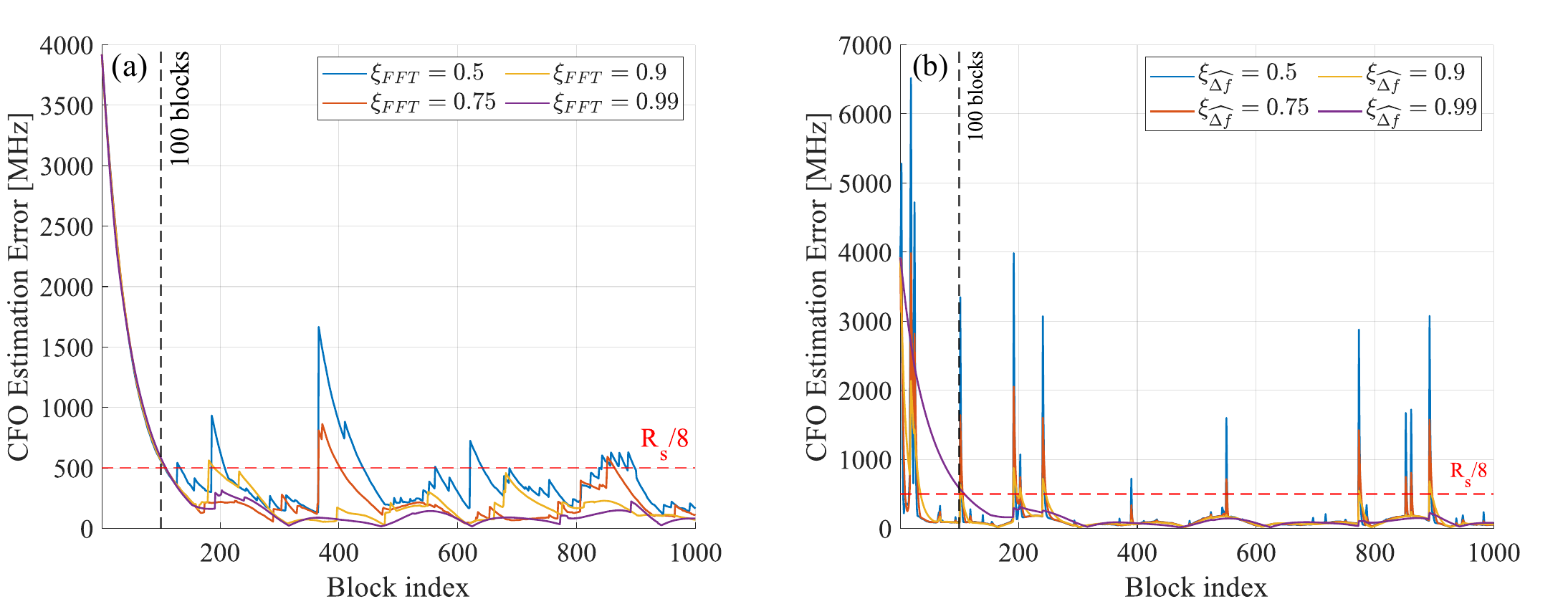}
    \caption{Impact of forgetting-factor selection on the convergence behavior of the proposed CFO estimator under the worst-case operating condition, with $R_s = 4~\mathrm{GBd}$, $\Delta f = 4~\mathrm{GHz}$, and $\mathrm{SNR}_b = 0~\mathrm{dB}$. The worst-case per-block estimation error over 50 channel realizations is shown as a function of the block index for different values of (a) the FFT forgetting factor, $\xi_{\mathrm{FFT}}$ (with $\xi_{\widehat{\Delta f}} = 0.99$), and (b) the CFO-estimate smoothing factor, $\xi_{\widehat{\Delta f}}$ (with $\xi_{\mathrm{FFT}} = 0.99$). The dashed vertical line marks the first 100 blocks discarded in the performance evaluation, while the red horizontal dashed line indicates the maximum tolerable CFO-estimation error for this transmission, defined by the 4th-power threshold.}
    \label{fig:forgetgect}
\end{figure}

Figure~\ref{fig:forgetgect} further analyzes the convergence behavior and the impact of the forgetting-factor selection under the most stringent operating condition considered in Fig.~\ref{fig:heatmap}, namely the $R_s = 4~\mathrm{GBd}$ case, with large CFO and low SNR. The results show that the forgetting factors control the trade-off between acquisition speed and post-convergence stability. In Fig.~\ref{fig:forgetgect}a, where $\xi_{\widehat{\Delta f}}$ is fixed at $0.99$, smaller values of $\xi_{\mathrm{FFT}}$ provide faster adaptation of the averaged spectrum, but also lead to larger fluctuations after the initial transient because less temporal averaging is applied to the PSD estimate. Conversely, values of $\xi_{\mathrm{FFT}}$ closer to unity increase the effective memory of the spectral estimate, reducing noise-induced fluctuations at the cost of a slower initial convergence. A similar behavior is observed in Fig.~\ref{fig:forgetgect}b, where $\xi_{\mathrm{FFT}}$ is fixed at 0.99 and $\xi_{\widehat{\Delta f}}$ is varied: lower values make the CFO estimator more responsive but, at the same time, more sensitive to outliers, whereas larger values smooth the estimate and improve tracking stability.

The vertical dashed line at 100 FFT blocks therefore represents a conservative convergence margin rather than an optimized value for a specific scenario. It is selected to ensure that the estimator reaches a stable operating region under demanding acquisition conditions, including the lowest evaluated symbol rate and SNR, together with a severe CFO. After this initial transient, the residual CFO remains predominantly below the $R_s/8$ threshold required by the subsequent F-CFOE stage. For the adopted FFT size of 1024 samples and sampling rate of $F_s = 64~\mathrm{GSa/s}$, this discarded interval corresponds to approximately $1.6~\mu\mathrm{s}$, which is short compared with the typical timescale of Doppler variations in LEO satellite links. These results indicate that forgetting factors close to unity provide a suitable compromise between convergence latency and tracking accuracy for the considered C-CFO estimator.

\begin{figure}[ht]
    \centering
    \includegraphics[width=0.65\linewidth]{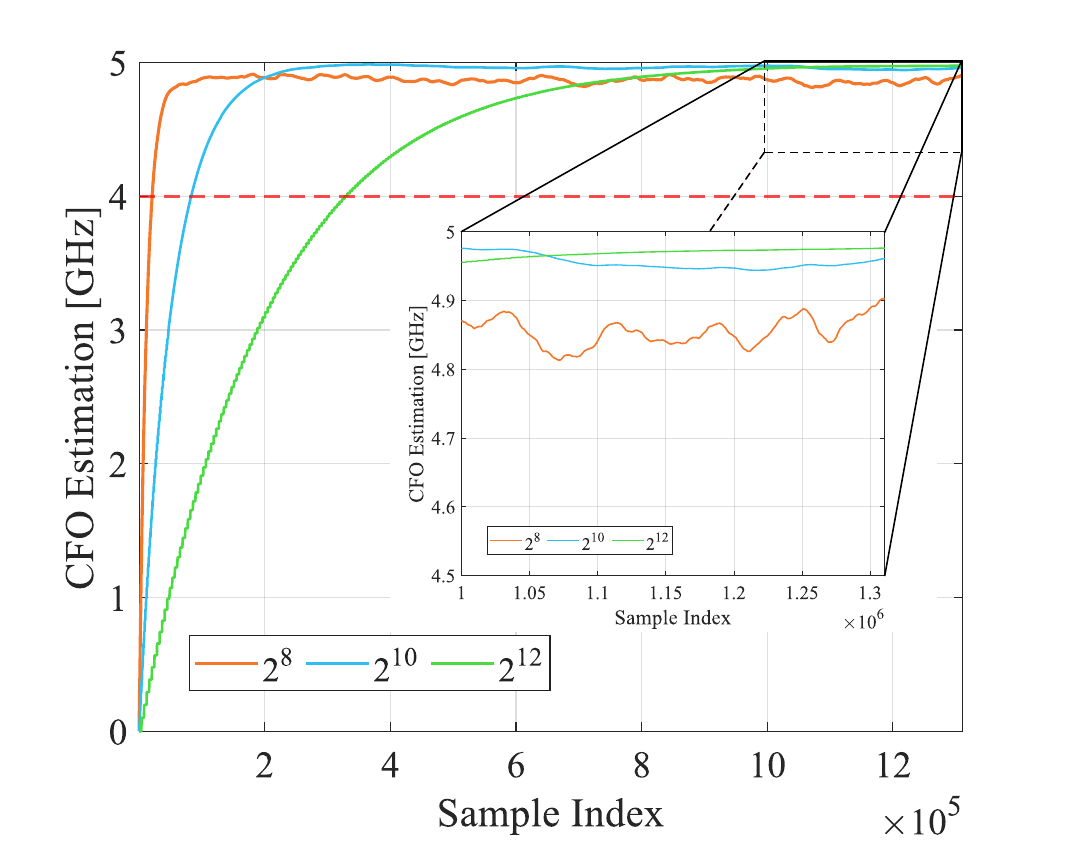}
    \caption{Impact of the FFT size on the convergence behavior of the proposed C-CFO estimator. The estimated CFO is shown as a function of the processed symbol index for FFT sizes of $2^8$, $2^{10}$, and $2^{12}$ samples, under a representative large-CFO condition of $5$~GHz, with $R_s = 8$~GBd and SNR$_b = 5$~dB. The forgetting factors are fixed at $\xi_{\mathrm{FFT}}=\xi_{\widehat{\Delta f}}=0.98$ for all cases. Smaller FFT sizes provide a faster initial response because each spectral estimate is obtained from a shorter observation window, but they exhibit stronger post-convergence fluctuations due to poorer frequency resolution and noisier PSD estimates. Larger FFT sizes improve the spectral resolution and stabilize the accumulated-PSD profile, at the expense of slower convergence in terms of processed symbols. The inset highlights the steady-state region within the $\pm R_s/8$ threshold (red dashed line) imposed by the subsequent 4th-power F-CFOE stage.}
    \label{fig:fftsize}
\end{figure}

A key trade-off in the implementation of the proposed estimator is governed by the FFT size, $N_{\mathrm{FFT}}$, which determines both the frequency resolution, $\delta f = F_s/N_{\mathrm{FFT}}$, and the observation window used to obtain the block-wise PSD estimate. Larger FFT sizes provide finer spectral resolution and a smoother accumulated-PSD profile, which can improve breakpoint localization. However, they also increase the number of processed symbols required for the estimator to react. In contrast, smaller FFT sizes reduce latency and accelerate the initial estimator response, but at the cost of increased spectral uncertainty and larger steady-state fluctuations. Figure~\ref{fig:fftsize} illustrates this trade-off for an $R_s = 8$~GBd transmission at $\mathrm{SNR}_b = 5$~dB under a constant CFO of 5~GHz. For the same forgetting factors, $\xi_{\mathrm{FFT}}=\xi_{\widehat{\Delta f}}=0.98$, the case $N_{\mathrm{FFT}}=2^8$ exhibits a faster initial response due to the shorter observation window, but the coarse spectral resolution leads to stronger post-convergence fluctuations. Conversely, $N_{\mathrm{FFT}}=2^{12}$ provides a more stable accumulated-PSD profile and reduced steady-state variation, but requires more processed symbols to reach convergence. Therefore, the adopted value, $N_{\mathrm{FFT}}=2^{10}$, represents a practical compromise between acquisition speed, spectral resolution, and steady-state estimation stability.

\subsection{Experimental Validation}

\begin{figure}[!bthp]
    \centering
    \includegraphics[width=0.95\linewidth]{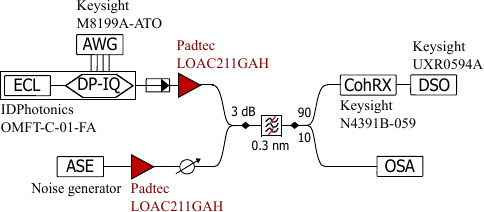}
    \caption{Experimental setup for the evaluation of the proposed C-CFOE algorithm. AWG: arbitrary waveform generator; DP-IQ: dual-polarization in-phase/quadrature modulator; VOA: variable optical attenuator; EDFA: erbium-doped fiber amplifier; OBPF: optical bandpass filter; OMA: optical modulation analyzer; DSO: digital storage oscilloscope.}
    \label{fig:setup}
\end{figure}

Experimental measurements were carried out to evaluate the C-CFOE performance using the setup depicted in Fig. \ref{fig:setup}. At the transmitter, digital waveforms are generated offline and replayed by an arbitrary waveform generator (AWG, Keysight M8199A-ATO) operating at 128~GSa/s. These waveforms correspond to a 15th and 11th-order pseudo-random binary sequence (PRBS-15 and PRBS-11, respectively, to create different sequences in each channel) transmitted at a symbol rate of 4~GBaud. This relatively low symbol rate was selected to emulate very stringent frequency recovery conditions where the frequency offset approaches or surpasses the signal bandwidth. The resulting electrical signals drive a dual-polarization in-phase/quadrature  modulator (DP-IQ, IDPhotonics OMFT-C-01-FA), which modulates a narrow linewidth ($<$25~kHz) continuous-wave optical carrier. An optical isolator is placed after modulation to suppress back-reflections and prevent optical feedback. After modulation, the optical signal is amplified by an erbium-doped fiber amplifier (EDFA, Padtec LOAC211GAH) and mixed with amplified spontaneous emission noise through a 3-dB coupler. The noise level is attenuated by a variable optical attenuator (VOA, Eigenlight 420 WDM) equipped with an integrated power meter to control noise power and consequently the SNR. The mixed signal and noise pass through a 37.5-GHz filter enabled by a wavelength selective switch (WSS) to ensure spectral consistency. This bandwidth is selected to ensure a flat spectrum around the signal and prevent cutting spectral content needed for DSP. The resulting shaped signal then passes through a 90:10 optical coupler: 90\% of the power is directed to the coherent receiver (CohRX), while the remaining 10\% is sent to an optical spectrum analyzer (OSA) with 0.1-nm resolution bandwidth for SNR measurement. Coherent detection is performed using a free-running local oscillator (LO) laser within an optical modulation analyzer (OMA, Keysight N4391B-059). The electrical outputs of the coherent receiver are subsequently sampled at 32~GSa/s by a high-speed digital storage oscilloscope (DSO, Keysight UXR0594A) and processed offline.

\begin{figure}[!htbp]
    \centering
   \includegraphics[width=0.65\linewidth]{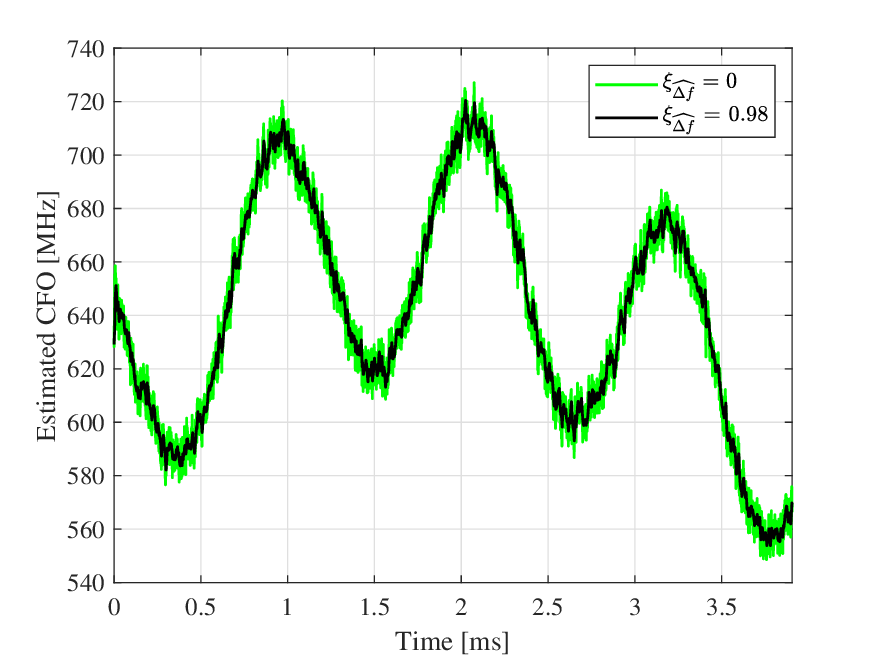}
    \caption{Experimental evolution of the estimator output for the unfiltered case (green) and the filtered case with a 0.98 forgetting factor (black). The frequency drift was estimated over a long-duration acquisition of approximately 4~ms, corresponding to 61,035 FFT blocks of 1,024 samples each, revealing a systematic offset of about 640 MHz between the transmitter and the LO.}
    \label{fig:4ms}
\end{figure}

Figure \ref{fig:4ms} shows the algorithm tracking over an acquisition time of $\sim$4~ms, at SNR$_b$ = 5~dB, corresponding to
61,035 FFT blocks of 1,024 samples each. The two curves indicate the unfiltered and filtered cases. The algorithm reveals a systematic frequency offset of about 640~MHz between the transmitter and the LO, along with peak-to-peak oscillations of about 150~MHz. As data acquisition was carried out at 4-GBd symbol rate, the obtained performance would not be achieved by the traditional 4th-power algorithm, which has an estimation bound of $R_s/8$, in this case, only 500 MHz.

To evaluate the capability of C-CFOE to reduce the CFO below the threshold required by the 4th-power algorithm, a series of experimental measurements was performed for a 4-GBd PM-QPSK transmission (16~Gbps). The SNR per bit was evaluated at three levels: 0, 5, and 10~dB. In addition to the intrinsic frequency offset between the Tx and LO lasers ($\sim$640 MHz, as previously evaluated), four additional scenarios were considered (-4~GHz, -2~GHz, 2~GHz, and 4~GHz), in which an intentional CFO was applied to the Tx laser. The data acquisition time for these evaluations was 500~$\mu$s.

\begin{figure}[!htbp]
    \centering
    \includegraphics[width=\linewidth]{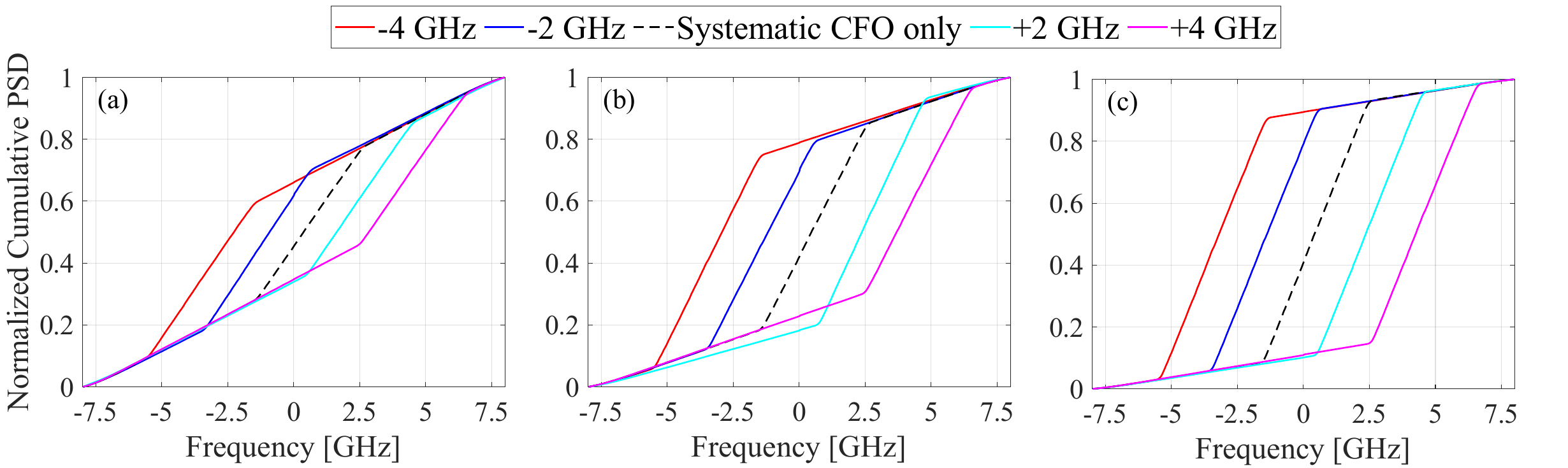}
    \caption{Experimental PSD accumulation curves relating to the last FFT block of each 4-GBd PM-QPSK transmission, at SNR per bit values of (a) 0 dB, (b) 5 dB and (c) 10 dB, under multiple carrier frequency offsets.}
    \label{fig:exp_slr}
\end{figure}

\begin{figure}[!htbp]
    \centering
    \includegraphics[width=\linewidth]{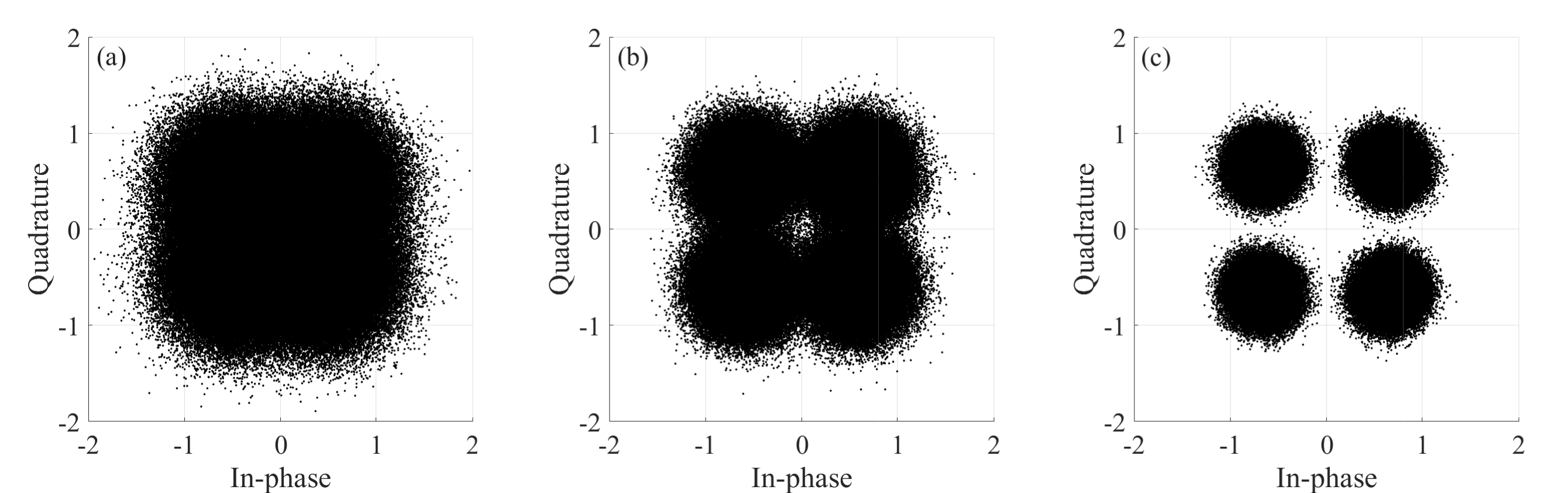}
    \caption{Recovered constellations for (a) $\mathrm{SNR}_b = 0~\mathrm{dB}$, (b) $\mathrm{SNR}_b = 5~\mathrm{dB}$, and (c) $\mathrm{SNR}_b = 10~\mathrm{dB}$, considering $R_s = 4~\mathrm{GBd}$ and $\Delta f = 4~\mathrm{GHz}$. The obtained BER values are $8.3895 \times 10^{-2}$, $6.7678 \times 10^{-3}$, and $7.9045 \times 10^{-6}$, respectively.
}
    \label{fig:constellation}
\end{figure}

Figure \ref{fig:exp_slr} shows the PSD accumulation profiles for the last processed blocks in each case considered. It can be observed that in the scenario where no external CFO is applied (dashed black curves), the midpoint of the central segment, referring to the signal, is slightly shifted to the right, in accordance with the systematic CFO shown in Fig. \ref{fig:4ms}. Consequently, all curves will result from the sum between the applied external CFO and the intrinsic one, even exceeding the signal bandwidth in the case of +4~GHz. As expected, Fig. \ref{fig:exp_slr}a, with lower SNR, results in less abrupt transitions compared to the slope changes present in Figs. \ref{fig:exp_slr}b or c, with higher SNR. Table \ref{tab:residualf-cfoe} summarizes the residual frequency offsets estimated using the 4th-power algorithm. The processing structure illustrated in Fig.~\ref{fig:diagram} was employed with $\xi_{\mathrm{FFT}} = \xi_{\widehat{\Delta f}} = 0.98$ for $\mathrm{SNR}_b = 0\,\mathrm{dB}$, and $\xi_{\mathrm{FFT}} = 0.98$ and $\xi_{\widehat{\Delta f}} = 0.95$ for all other cases. Based on the characterization presented in Fig. \ref{fig:4ms}, $\Delta f_{\max}$ was set to 5~GHz. The C-CFOE stage is applied after the Gram–Schmidt orthonormalization process (GSOP) and DC offset removal. Following C-CFOE, adaptive equalization is performed using the constant-modulus algorithm (CMA) with single-spike initialization (15,000 samples for initialization), 11 taps, and a step size of $5\cdot10^{-3}$. Once the signal is equalized and resampled to 1 sample per symbol, the 4th-power method is applied. The results reveal an excellent C-CFOE performance, generating estimates that are well below the 500-MHz correction limit imposed by the F-CFOE M-th power algorithm. In all cases the recovered constellations after phase recovery demonstrate a noisy QPSK aspect, without indication of possible residual frequency offsets after F-CFOE, as illustrated in Fig.~\ref{fig:constellation} for the cases in which a 4-GHz CFO was applied to the Tx laser.

\begin{table}[!htbp]
\centering
\caption{Experimental maximum residual CFO estimation error per polarization, in MHz, obtained after C-CFOR using the proposed estimator and evaluated with the 4th-power method. During the coarse estimation stage, the first 100 FFT blocks were used for convergence and subsequently discarded.}
\label{tab:residualf-cfoe}
\begin{tabular}{c || c c c}
\hline
\textbf{Scenario} & 0 dB & 5 dB & 10 dB \\
\hline\hline
-4 GHz & 110.592, 155.9468  & 16.6071, 19.2609 &  17.9999, 17.5546\\
-2 GHz & 79.2561, 87.3008   & 31.6661, 31.146 & 4.3226, 5.5959 \\
0 & 63.6535, 63.4917  & 21.1226, 24.5414 & 12.6113, 12.042 \\
+2 GHz & 114.7825, 120.7017 & 37.4987, 39.202 & 24.1993, 24.3575 \\
+4 GHz & 227.755, 262.2029 & 31.0037, 36.1317 & 15.0848, 15.3057 \\
\hline
\end{tabular}
\end{table}

\FloatBarrier

\section{Conclusion}
\label{sec:conclusion}

We propose a C-CFOE algorithm for stringent operating conditions in coherent optical satellite communications, particularly in LEO-to-Earth distance-adaptive links, where low SNR, low symbol rate, and large frequency offsets may occur simultaneously. In such scenarios, the CFO can approach or exceed the signal bandwidth, while conventional fine CFOE techniques, such as those based on the $M$-th power method, are limited to a fraction of the symbol rate and therefore require a reliable coarse acquisition stage.\\
\indent The proposed method exploits the fact that the accumulated PSD of a Nyquist-shaped signal received with excess bandwidth exhibits an approximately three-segment piecewise-linear structure. In contrast to spectral methods based on local features, such as individual spectral edges, asymmetry metrics, or power differences over predefined frequency regions, the proposed estimator uses the global shape of the accumulated spectrum. The transition points between the noise-floor and signal-occupied regions are identified through a low-complexity segmented linear regression procedure, from which the coarse CFO estimate is obtained.\\
\indent Simulation and experimental results demonstrate that the proposed estimator can robustly reduce large CFOs to residual values compatible with the acquisition range of subsequent fine CFOE algorithms, even under low-symbol-rate and low-SNR conditions. The method is therefore suitable as a coarse synchronization stage preceding conventional fine carrier recovery. Its main performance limitation arises when the accumulated PSD no longer preserves sufficiently distinguishable piecewise-linear regimes. Overall, the proposed SLR-based C-CFOE provides a low-complexity solution for wide-range CFO acquisition in coherent optical satellite links.

\appendix

\section{Linearization via Integral Equations and SLR}
\label{appA}

Here, we introduce the linearization procedure using integral equations, originally developed by Jean Jacquelin \cite{Jacquelin2009}, and summarize the closed-form analytical expressions employed for the SLR-based CFOE procedure. For clarity, the presentation in this Appendix is restricted to the linear case. Nevertheless, the methodology is general and can be extended to arbitrary functional forms.

The function to be fitted is defined piecewise as
\begin{equation}
y(x)\equiv
\begin{cases}
p_1x+q_1, & x\in(-\infty,\psi_1),\\[3pt]
p_2x+q_2, & x\in[\psi_1,\psi_2),\\[3pt]
p_3x+q_3, & x\in[\psi_2,\infty),
\end{cases}
\end{equation}
with $\psi_1<\psi_2$ assumed without loss of generality. An equivalent representation can be obtained using the Heaviside step function $H\left(\cdot\right)$:
\begin{equation}
\begin{aligned}
y(x) ={}&(p_1x+q_1)\bigl[1-H(x-\psi_1)\bigr]+(p_2x+q_2)\bigl[H(x-\psi_1)-H(x-\psi_2)\bigr]\\
&+(p_3x+q_3)H(x-\psi_2).
\nonumber
\end{aligned}
\end{equation}
The function $y(x)$ can then be rewritten in a form that explicitly captures the discontinuities of its first derivative at the breakpoints, namely
\begin{equation}
    y(x) = p_1x+q_1+(p_2-p_1)(x-\psi_1)H(x-\psi_1)+(p_3-p_2)(x-\psi_2)H(x-\psi_2),
    \label{eq:y}
\end{equation}
from which the following system of equations is obtained by integration:
\begin{equation}
\left\{
\begin{aligned}
y(x) &= p_1x+q_1
      + \sum_{i=1}^2 (p_{i+1}-p_i)(x-\psi_i)H(x-\psi_i),\\[6pt]
2\int^{x} y(s)\,ds
&= p_1x^2+2q_1x
      + \sum_{i=1}^2 (p_{i+1}-p_i)(x-\psi_i)^2H(x-\psi_i)
      + c_0,\\[6pt]
6\int^{x}\!\!\int^{s} y(t)\,dt\,ds
&= p_1x^3+3q_1x^2
      + \sum_{i=1}^2 (p_{i+1}-p_i)(x-\psi_i)^3H(x-\psi_i)
      + c_1x+c_2.
\end{aligned}
\right.
\nonumber
\end{equation}
where $c_0$, $c_1$, and $c_2$ are arbitrary constants of integration.

By eliminating the Heaviside-dependent components $(p_2-p_1)(x-\psi_1)H(x-\psi_1)$ and $(p_3-p_2)(x-\psi_2)H(x-\psi_2)$ from the system and rearranging the remaining terms, we obtain the following integral equation:
\begin{equation}
\begin{aligned}
y(x)
={}&\frac{1}{\psi_1\psi_2}\left(
4x\int^{x} y(s)\,ds
-6\int^{x}\!\!\int^{s} y(t)\,dt\,ds
-x^2y(x)\right) \\
&+\frac{\psi_1+\psi_2}{\psi_1\psi_2}\left(
xy(x)-2\int^{x} y(s)\,ds\right)+c_3x+c_4.
\end{aligned}
\end{equation}
Using the integration-by-parts identity
\begin{equation}
\int^x\!\!\int^s y(t)\,dt\,ds
= x\int^x y(s)\,ds - \int^x s\,y(s)\,ds + C_3 x + C_4,
\nonumber
\end{equation}
it then follows that
\begin{equation}
\begin{aligned}
y(x)
={}& \frac{1}{\psi_1\psi_2}\!\left(
6\int^{x} s\,y(s)\,ds -2x\int^{x} y(s)\,ds
-x^2y(x)\right)+\frac{\psi_1+\psi_2}{\psi_1\psi_2}\!\left(
x y(x)-2\int^{x} y(s)\,ds\right)\\
& + C_3 x + C_4.
\label{eq:int}
\end{aligned}
\end{equation}
The constants $C_3$ and $C_4$ represent homogeneous contributions and play no role in determining the breakpoints. Introducing the coefficients
\begin{equation}
C_1 \equiv \frac{1}{\psi_1 \psi_2}
\qquad
\textrm{ and }
\qquad
C_2 \equiv \frac{\psi_1 + \psi_2}{\psi_1 \psi_2},
\nonumber
\end{equation}
Eq. \eqref{eq:int} becomes linear with respect to $(C_1,C_2,C_3,C_4)$.

For discrete data $\{(x_k,y_k)\}_{k=1}^{N}$, the integral terms are now approximated by trapezoidal rule:
\begin{align}
S_y[1] &= 0, &
S_y[k] &= S_y[k-1] + \frac{1}{2}(y_{k-1}+y_k)(x_k-x_{k-1}), \\
S_{xy}[1] &= 0, &
S_{xy}[k] &= S_{xy}[k-1]
+ \frac{1}{2}(x_{k-1}y_{k-1}+x_k y_k)(x_k-x_{k-1}),
\quad 2 \le k \le N.
\end{align}

Evaluating \eqref{eq:int} at $x=x_k$ yields the linear regression model
\[
F_{0,k} = C_1 F_{1,k} + C_2 F_{2,k} + C_3 F_{3,k} + C_4 F_{4,k},
\]
where
\[
\begin{aligned}
F_{0,k} &= y_k, \quad
F_{1,k} = 6S_{xy}[k] - 2x_k S_y[k] - x_k^{2} y_k, \\
F_{2,k} &= x_k y_k - 2S_y[k], \quad
F_{3,k} = x_k, \quad
F_{4,k} = 1.
\end{aligned}
\]

The coefficients $\left\{C_i\right\}_{i=1}^4$ are obtained by least squares:
\begin{equation}
(C_1,C_2,C_3,C_4)^{\mathsf T} = \arg\min_{\mathbf c \in \mathbb{R}^4}\sum_{k=1}^{N}\left(F_{0,k} - c_1 F_{1,k} - c_2 F_{2,k} - c_3 F_{3,k} - c_4 F_{4,k}\right)^2.
\end{equation}
From $C_1$ and $C_2$, the breakpoints are recovered as the roots of the quadratic equation $C_1 t^2 - C_2 t + 1 = 0$:
\begin{equation}
\psi_{1,2} = \frac{C_2 \mp \sqrt{C_2^2 - 4C_1}}{2C_1},
\qquad \psi_1 < \psi_2.
\end{equation}

Once $\psi_1$ and $\psi_2$ are determined, they are substituted into \eqref{eq:y}, which becomes linear with respect to $(p_1,p_2,p_3,q_1)$. A second linear least-squares regression then yields the slopes of the three segments and the intercept $q_1$ of the first segment. The remaining intercepts follow directly from continuity at the breakpoints:
\begin{align}
    q_2 &= (p_1 - p_2)\psi_1 + q_1, \\
    q_3 &= (p_2 - p_3)\psi_2 + q_2.
\end{align}

\clearpage

\bibliography{optica}
	
\end{document}